\documentclass[journal]{IEEEtran}
\usepackage[pdftex]{graphicx}
\usepackage{epstopdf}
\usepackage{amsmath}
\usepackage{lineno}
\usepackage{amssymb}
\usepackage{subfigure}
\usepackage{booktabs}
\usepackage{cite}
\usepackage{color}
\usepackage[T1]{fontenc}
\usepackage{bm}
\usepackage{geometry}
\usepackage{tabularx}

\ifCLASSINFOpdf
\else
\fi
\begin{document}
\title{Distributed Event-Triggered Nonlinear Fusion Estimation under Resource Constraints}

\author{Rusheng~Wang,
        Bo~Chen,~
        Zhongyao~Hu,
        and Li~Yu% <-this % stops a space
%\thanks{This work was supported in part by the National Natural Science Foundation of
%China under Grant 61973277, Grant 62073292, and in part by the National Natural Science
%Foundation of China, and in part by the Zhejiang
%Provincial Natural Science Foundation of China under Grant LR20F030004.}% <-this % stops a space
\thanks{R. Wang, B. Chen, Z. Hu, L. Yu are with Institute of Cyberspace Security, Zhejiang University of Technology, Hangzhou 310023, China, and also with Department of Automation, Zhejiang University of Technology, Hangzhou 310023, China (email: wrsheng\_12@163.com, bchen@aliyun.com, huzhongyao@aliyun.com, lyu@zjut.edu.cn).}% <-this % stops a space
\thanks{Corresponding author: Bo Chen.}}

%\markboth{IEEE Transactions on Signal Processing}%

\maketitle

\begin{abstract}
 This paper studies the event-triggered distributed fusion estimation problems for a class of nonlinear networked multi-sensor fusion systems without noise statistical characteristics. When considering the limited resource problems of two kinds of communication channels (i.e., sensor-to-remote estimator channel and smart sensor-to-fusion center channel), an event-triggered strategy and a dimensionality reduction strategy are introduced in a unified networked framework to lighten the communication burden. Then, two kinds of compensation strategies in terms of a unified model are designed to restructure the untransmitted information, and the local/fusion estimators are proposed based on the compensation information. Furthermore, the linearization errors caused by the Taylor expansion are modeled by the state-dependent matrices with uncertain parameters when establishing estimation error systems, and then different robust recursive optimization problems are constructed to determine the estimator gains and the fusion criteria. Meanwhile, the stability conditions are also proposed such that the square errors of the designed nonlinear estimators are bounded. Finally, a vehicle localization system is employed to demonstrate the effectiveness and advantages of the proposed methods.
\end{abstract}


% \fi
\IEEEpeerreviewmaketitle

\section{Introduction}
\IEEEPARstart{T}{o} meet the requirements of higher estimation accuracy, various multi-sensor fusion estimation methods have been developed to improve the reliability and estimation performance by leveraging the redundant information \cite{r1,r2,r3}. Subsequently, with the development of communication networks and sensor technologies, networked multi-sensor fusion systems (NMFSs) have attracted extensive attention, in which the communication between the sensors, estimators, and fusion center (FC) is connected by the network instead of a dedicated independent connection \cite{m1}. Since the introduction of the communication networks provides a more flexible structure and easier installation and maintenance than the multi-sensor fusion systems, NMFSs have been applied in many areas, such as fault detection \cite{b1}, target tracking \cite{b30}, and power systems \cite{t1}. In particular, nonlinear NMFSs have complex dynamic processes that are more suitable for practical applications.  Moreover, the distributed fusion framework of nonlinear NMFSs has certain advantages in reliability, fault tolerance, and calculation speed \cite{b7}. In this case, some nonlinear distributed network fusion estimation problems caused by the introduction of the network will be considered in this paper.

Though the introduction of the network has brought a lot of convenience, it has also inevitably been confronted with some problems. Particularly, the communication channel can only carry finite information per unit time, and the communication capability is also restricted by the limited energy of the sensor nodes \cite{b6}.
Then, loss of information, which caused by the above resource constraints, can lead to a degradation of estimation performance. Therefore, it is of great significance to actively reduce the communication traffic to meet the limited communication resources, while a relatively satisfactory estimation performance of the designed estimator should also be guaranteed. Generally, various quantization strategies \cite{qu1,qu2,qu3} and dimensionality reduction strategies \cite{dr0,dr1,dr11,tsp1,q21,dr3,dr4,dr2,tsp9,qd1} have been developed to reduce the size of data packets before information transmission. For example, the optimal quantization rules and an optimal fusion estimation criterion were established in \cite{qu1}, and a logarithmic quantizer was utilized in \cite{qu2} to address the $H_\infty$ fusion estimation problem with restricted bandwidth. However, the quantization technique is difficult to deal with high-dimensional signals and has the limitation of being easily distorted. At the same time, the event-triggered transmission strategies \cite{b9,b17,tsp2,tsp3,tsp4,tsp5,tsp6,tsp61,b29,b31,tsp7,tsp8,tsp10,tsp11,tsp12,b11,b32,b18,b311} also provide an effective means to reduce redundant communication traffic and energy resources. In this case, a kind of dimensionality reduction strategy (DRS) and an event-triggered strategy (ETS) will be considered in this paper to deal with the networked fusion estimation problems under resource constraints.

It should be noted that limited communication resources may occur between the sensor and the estimator, or between the sensor and the FC. When considering that the measurements need to be sent to the remote estimator for processing, the limited resources problem of the sensor-to-remote estimator (S-RE) communication channel should be addressed. Therefore, an optimal compression matrix can be found in \cite{dr1} by an optimal sensor compression strategy. In particular, a dimensionality reduction strategy, which needs to know the global measurement matrix, has been developed in \cite{dr11} to compress the measurements. Meanwhile, a preprocessor was designed in \cite{tsp1} for compressing the raw measurements of the corresponding measurement block, while the contraction operators were usually difficult to solve. Then, another dimensionality reduction approach, in which only partial components of a group of measurements were chosen to be transmitted to the remote estimator, was proposed in \cite{q21} to meet the limited bandwidth. However, the approaches mentioned above are confined to linear NMFSs with Gaussian noises assumption.

In addition, the transmission of information in event-based networked system depends on predefined event-triggered conditions \cite{b9}. Specifically, a measurement innovation-based deterministic ETS was designed in \cite{tsp2}, and then a more general ETS was considered in \cite{tsp3} to address the fusion estimation problem based on the hybrid measurement information, while the conditional distribution of state was assumed to be a Gaussian distribution. Moreover, a kind of stochastic ETS was developed in \cite{tsp4,tsp5}, which can preserve the Gaussian property of the innovation sequence and the posterior distribution. However, the above innovation-based ETSs need feedback information from the estimator, which might increase communication costs and be difficult to apply to time-varying systems. Then, an information-based stochastic ETS based on the local observation projection into the state-space was developed in \cite{tsp6} instead of combine the feedback from the estimator, while it still considered the fusion estimation problem in the linear multi-sensor fusion framework. For the nonlinear systems, a measurement-based ETS was introduced in \cite{tsp61} to improve the utilization of network resources, and the T-S fuzzy technology was employed to cope with the nonlinearities. Then, a kind of distributed measurement innovation-based ETS was proposed in \cite{b29} to cut down on bandwidth usage, and then a distributed recursive estimation algorithm was designed under the stochastic nonlinearities and measurement losses. In \cite{b31}, the innovation-based deterministic ETS was used for nonlinear NMFSs with random delays, and then a modified unscented Kalman filter (UKF) and sequential covariance intersection (SCI) fusion method were proposed to address the distributed fusion estimation problem. Meanwhile, a stochastic Send-on-Delta (SoD) ETS was introduced in \cite{tsp7} to reduce the redundant information for nonlinear NMFSs subject to jamming attacks. Notice that, the above methods are considered in an estimation framework in which the measurements and/or noises have a Gaussian assumption with known covariance. Though a dynamic ETS was developed in \cite{tsp8} for nonlinear/non-Gaussian distributed system, it still requires knowledge of the nonlinear functions and noise probability density functions of the systems.

On the other hand, some smart sensors have the ability to process data locally instead of sending it to remote sensors, and then local estimates will be further transmitted to the FC through the communication network in the NMFSs. Thus, the resource constraints problem of the smart sensor-to-fusion center (S-FC) channel also should be considered. Then, a fixed structure DRS, in which only partial components of local estimates with known bandwidth constraints were chosen to transmit to the FC, was developed in \cite{dr3} to meet the limited bandwidth, and a suboptimal transmission sequence was also given. By using this DRS, a security fusion estimation problem under denial-of-service (DoS) attacks and resource constraints was addressed in \cite{dr4}. Moreover, another kind of DRS without a fixed structure was developed in \cite{dr2}, in which the partial local estimates were randomly selected to be sent, and a compensation strategy was proposed to restructure the untransmitted information. Notice that the above methods are all based on the Kalman filter, which needs to know the noise statistical characteristics. Then, a $H_\infty$ fusion estimation method was developed in \cite{tsp9} to deal with finite communication resources, while the energy noise was still a kind of special noises. Though a DRS and quantization strategy (QS) were modeled by a unified framework with unknown bounded noises in \cite{qd1}, the considered problem was still in the linear distributed fusion framework.

Furthermore, a Gaussianity-preserving ETS was proposed in \cite{tsp10} the incremental innovative information of the estimates. Then, a deterministic ETS and a stochastic ETS, whose triggering conditions were based on the local estimates, were introduced in \cite{tsp11} and \cite{tsp12} for distributed state estimation problems, respectively.
Moreover, a new variance-based ETS was developed in \cite{b11} for the distributed estimation problem, in which the triggering condition was based on the difference between the estimation error variance and the multistep prediction variance. Particularly, a priori estimate-based ETS was proposed in \cite{b32} for nonlinear system without knowledge of process noise statistical property, while the covariances of measurement noise were still required to be known. Notice that, the above event-based estimation methods still consider the resource constraints problem of the S-RE channel rather than the S-FC channel of nonlinear NMFSs. In this case, though a deterministic ETS combined with DRS was proposed in \cite{b18} to address the distributed fusion estimation problem under resource constraints, it was still only suitable for linear NMFSs with Gaussian white noise. Then, a fusion estimate-based ETS was developed in \cite{b311}, in which the raw measurements and the feedback fusion information were employed to derive the local estimation, whereas it might increase the computation cost because of this fusion feedback.

In light the analysis mentioned above, there are abundant works have been developed to address the networked state estimation problems under resource constraints. However, still few works have paid attention to nonlinear fusion estimation problems for nonlinear NMFSs with unknown noise statistical property. Under this case, we shall study the event-based nonlinear distributed fusion estimation problems under resource constraints, where the noises are without knowledge of statistical property. The main contributions of this paper are summarized as follows:
\begin{enumerate}
\item In this paper, two different means of communication transmission are presented in the nonlinear networked fusion estimation framework. Specifically, both S-RE channel and S-FC channel subject to resource constraints are considered, and then a deterministic ETS and a DRS are both introduced in a unified framework to meet the finite resources.
\item Two kinds of unified compensation strategies are proposed to restructure the missed information caused by the ETS and the DRS, and then the corresponding distributed fusion estimation algorithms are developed, which can preserve a satisfying estimation performance.
\item By modeling the linearization errors in terms of the state-dependent matrices and uncertain parameters, a robust recursive optimization approach, which can deal with the unknown but bounded noises in nonlinear NMFSs, is developed such that the square errors (SEs) of the designed nonlinear local/fusion estimators are bounded.
\end{enumerate}

The notations most frequently used throughout this paper are given in TABLE I.
\begin{table}[htbp]
\centering
\caption{The description of the notations}
\begin{tabularx}{8.8cm}{ll}
\toprule
Notations & Descriptions \\
\midrule
$\mathfrak{R}^{m\times n}$ & Set of $m\times n$ real matrices \\
$\mathfrak{N}_+$ & Set of positive integer \\
$A^{\mathrm{T}}$ & Transpose of matrix $A$ \\
$A<0$ & Negative definite matrix \\
$I$ & Identity matrix with appropriate dimension \\
$``*"$ & Symmetric term of the symmetric matrix \\
$\|\cdot\|_2$ & 2-norm of the matrix\\
$\mathbb{E}\{\cdot\}$ & Mathematical expectation\\
$\mathrm{Tr}\{\cdot\}$ & Trace of the matrix \\
$\mathrm{col}\left\{\cdot \right\}$ & Block column matrix\\
$\mathrm{{diag}}\left\{\cdot \right\}$ & Block diagonal matrix \\
$\lambda_{\max}(\cdot)$ & Maximum eigenvalue of the matrix \\
[-0.5mm]
\bottomrule
\end{tabularx}
\end{table}
\section{Problem Formulation}
Consider a NMFS with $L$ sensor nodes, and the nonlinear state-space model can be modeled by:
\begin{align}\label{eq1}
	&{x}(k+1) = {f}({x}(k)) + {\Gamma}(k){w}(k) \\
	\begin{split}\label{eq2}
   &{{y}_i}(k) = {h_i}(x(k)) + {{D}_i}(k){{v}_i}(k), ~i \in {\mathfrak{L}}
	\end{split}
\end{align}
where $ x(k) \in {\mathfrak{R}^n}$, ${y_i}(k)\in \mathfrak{R}^{m_i}$ are the system state and measured output of $i$-th sensor, respectively, and $\mathfrak{L} \triangleq \{1,2,\ldots,L\}$.
${f}(x(k)) \in {\mathfrak{R}^{n}}$ and ${h_i}(x(k))\in \mathfrak{R}^{m_i}$ are continuously differentiable nonlinear functions, ${\Gamma}(k)\in {\mathfrak{R}^{n \times p}}$ and ${D}_i(k) \in {\mathfrak{R}^{{m_i}\times {q_i}}}$ are time-varying matrices.
${w}(k) \in {\mathfrak{R}^p}$ and ${{v}_i}(k) \in {\mathfrak{R}^{q_i}}$ are bounded noises without statistical information, which satisfy the following assumptions:
\begin{equation}\label{eq3}
\|w(k)\|_2 \le {\mathcal{M}_{w}},\;\; \|v_i(k)\|_2  \le {\mathcal{M}_{v_i}}
\end{equation}
where $\mathcal{M}_{w}$ and $\mathcal{M}_{v_i}$ are unknown. Generally, when designing the estimator based on the raw measurement information $\{y_i(1),y_i(2),\ldots,y_i(k)\}$, i.e. the network resource constraints are not considered. Then, the local nonlinear estimator (LNE) can be given by \cite{q2}:
\begin{eqnarray}\label{eq4}
\left\{
\begin{aligned}
&\hat{x}^-_i(k)=f(\hat{x}^s_i(k-1))\\
&\hat{x}^s_i(k)=\hat{x}^-_i(k)+{\mathrm{K}}^s_i(k)[y_i(k)-h_i(\hat{x}^-_i(k))]
\end{aligned}
\right.
\end{eqnarray}
where $\hat{x}^-_i(k)$ and $\hat{x}^s_i(k)$ are one-step prediction and local estimate, respectively,  $\mathrm{K}^{s}_i(k)$ is the estimator gain matrix. Under this case, in order to minimize the upper bound of the estimation error, the $\mathrm{K}^s_i(k)$ and $\hat{x}^s_i(k)$ can be obtained by using a robust design approach \cite[Th.1]{q2}, the detailed deduction is omitted here.

\subsection{Event-Triggered Strategy and Dimensionality Reduction Strategy over S-RE Channel}
In networked multi-sensor fusion systems, the communication network allocating enough bits to transfer the raw measurements from the sensor to the remote estimator is usually impracticable, then the limitation of the network resources in S-RE channel should be taken into account when designing nonlinear estimators. In this case, an event-triggered strategy (ETS) is introduced to alleviate the network resource burden. Specifically, a pre-designed event-triggered condition is given to calculate the decision variable $\gamma_m^i(k)(\in \{0,1\})$, which decides whether the raw measurement ${y}_i(k)$ is transmitted to the corresponding remote estimator or not. Then, an event-triggered condition for the $i$-th S-RE channel is given as
\begin{eqnarray}\label{ea12}
&\gamma_m^i(k)=\left\{
\begin{aligned}
&0, \quad \mathrm{if} \; y_i(k)\in \mathfrak{D}^i_m(k)\\
&1, \quad \mathrm{otherwise}
\end{aligned}
\right.  \\
&\mathfrak{D}^i_m(k)=\left\{\begin{aligned}y_i(k)\left|\right.\|{y}_i(k)-\breve{y}_i(k)\|_2 \leq \delta_m^i, i\in \mathfrak{L} \end{aligned}\right\} \label{eaa12}
\end{eqnarray}
where $\delta_m^i > 0$ is a predetermined triggering threshold, $\breve{y}_i(k)$ is the measurement for further dimensionality reduction processing at the latest event instant, $\mathfrak{D}^i_m(k)$ is a measurement set of each sensor that the event is not triggered at that time. Then, it can be found from the ETS \eqref{ea12}-\eqref{eaa12} that the raw measurement ${y}_i(k)$ will be further processed when $\gamma_m^i(k)=1$, otherwise there is no measurement information sent to the remote estimator at time $k$. Thus, the event-triggered instants sequence $0\leq t^i_m(1)\leq \cdots \leq t^i_m(k) \leq \cdots$ is determined by
\begin{equation}\label{ere1}
  t^i_m(k+1)=\min \left\{\begin{aligned} k \in \mathfrak{N}_+ \left|\right. k>t^i_m(k),\; y_i(k) \notin \mathfrak{D}^i_m(k) \end{aligned}\right\}
\end{equation}

On the other hand, in order to further reduce redundant communication traffic, a kind of dimensionality reduction strategy (DRS) \cite{dr3} is also introduced to reduce the size of the measurements from the event-triggered scheduler. Under this DRS, only $\varsigma^i_m$ components of measurement $\gamma^i_m(k)y_i(k)$ are allowed to be sent to the remote estimator at each time, rather than all information of $\gamma^i_m(k)y_i(k)$. Notice that, the measurement information does not need to be dimensionally reduced at a certain non-triggered moment (i.e. $\gamma^i_m(k)=0$). Moreover, it is assumed that the global bandwidth constraint is known advance:
\begin{equation}\label{re1}
  \sum_{i=1}^{L}\varsigma^i_m=\varsigma_m \;\left(\varsigma_m \in \mathfrak{N}_+,\varsigma^i_m \in \mathfrak{N}_+, 1\leq \varsigma^i_m \leq m_i\right)
\end{equation}
where $\varsigma_m$ is known. Then, the remote estimator may receive one of the $\mathfrak{h}_m^i$ types of dimensionality reduction measurements from the corresponding sensor through the network, where $\mathfrak{h}_m^i=\prod^{\varsigma^i_m-1}_{\imath=0}(m_i-\imath)/\prod_{\jmath=1}^{\varsigma^i_m}\jmath$. In other words, only one dimensionality reduction information from the following set will be transmitted at each instant:
\begin{equation}\label{ea2}
{\boldsymbol{\chi}}_m^i(k) \triangleq \left\{\gamma^i_m(k){\Theta}^i_{\kappa_i}y_i(k)\left|\right.\kappa_i\in {\mathfrak{H}}^i_m \triangleq \left\{1,2,\ldots,\mathfrak{h}_m^i\right\}\right\}
\end{equation}
where $\Theta^i_{\kappa_i}$ is a 0-1 diagonal matrix, whose diagonal terms including $\varsigma_m^i$ elements ``1''. In fact, matrix $\Theta^i_{\kappa_i}$ shows the dimensionality reduction status of each measurement. Under this case, a binary variable $\sigma^i_{\kappa_i}(k)(\in \{0,1\})$ is employed to describe the dimensionality reduction matrix ${\Theta}_m^i(k)$ in a clear way, and then the final transmission measurement $z_i(k)$ of each sensor can be modeled by:
\begin{equation}\label{ea5}
z_i(k)=\gamma^i_m(k)\sum^{\mathfrak{h}_m^i}_{\kappa_i=1}\sigma^i_{\kappa_i}(k){\Theta}^i_{\kappa_i}y_i(k)=\gamma^i_m(k){\Theta}_m^i(k)y_i(k)
\end{equation}
where $\sum_{\kappa_i=1}^{\mathfrak{h}_m^i}\sigma^i_{\kappa_i}(k)=1$, and it can be seen from ${\Theta}_m^i(k)=\sum^{\mathfrak{h}_m^i}_{\kappa_i=1}\sigma^i_{\kappa_i}(k){\Theta}^i_{\kappa_i}$ that dimension reduction matrix ${\Theta}_m^i(k)$ is decided by $\sigma^i_{\kappa_i}(k)$.

\begin{figure}[!htb]
\begin{center}
  \setlength{\abovecaptionskip}{0.1cm}
  \centering
  \includegraphics[width=9cm, height=5cm]{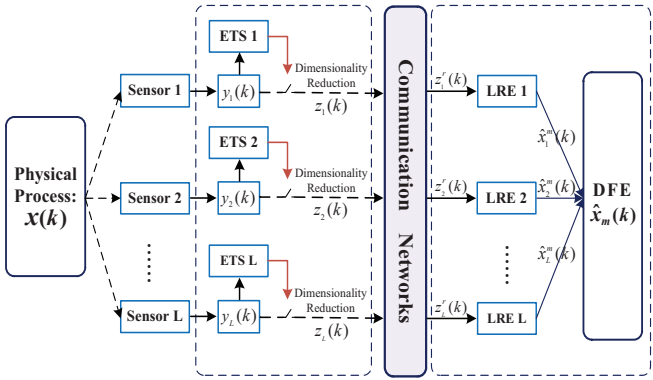}
\caption{The diagram of the modeling process when the S-RE channel is subject to resource constraints.}
  \label{fig1}
  \end{center}
\end{figure}

Since an ETS and a DRS are both employed to address the problem of networked fusion estimation under resource constraints, the estimation performance will be inevitably degraded. To ensure the estimator has better estimation performance, a unified compensation model is proposed in this paper to compensate the untransmitted measurement information. Then, the following compensation measurement can be used for the remote estimator at each time:
\begin{equation} \label{ea14}
{z}^r_i(k)={z}_i(k)+(I-\gamma_m^i(k)\Theta_m^i(k))h_i(\hat{x}^{-}_{m_i}(k))
\end{equation}
It can be concluded from the above compensation model that the prediction measurement $h_i(\hat{x}^{-}_{m_i}(k))$ is employed to estimate the system state when there is no measurement received by the remote estimator (i.e. $\gamma_i(k)=0$); otherwise the $(I-\gamma_m^i(k)\Theta_m^i(k))h_i(\hat{x}^{-}_{m_i}(k))$ is used to compensate the dimensionality reduction components of the measurement $y_i(k)$ when $\gamma_i(k)=1$. The structure of distributed fusion estimation under resource constraints based on the above modeling process is depicted in Fig. \ref{fig1}.

Based on the above communication strategy under resource constraints, the event-triggered local remote estimator (LRE) for the systems (\ref{eq1}) and (\ref{ea14}) is given by
\begin{eqnarray}\label{eq8}
\left\{
\begin{aligned}
&\hat{x}^{-}_{m_i}(k)=f(\hat{x}^m_i(k-1))\\
&\hat{x}^m_i(k)=\hat{x}^{-}_{m_i}(k)+{\mathrm{K}}^m_i(k)[{z}^r_i(k)-h_i(\hat{x}^{-}_{m_i}(k))]
\end{aligned}
\right.
\end{eqnarray}
where $\hat{x}^{-}_{m_i}(k)$ and $\hat{x}^m_i(k)$ are one-step prediction and local estimate, respectively, ${\mathrm{K}}^m_i(k)$ is the estimator gain need to be designed.

Then, on the basis of the LRE \eqref{eq8}, the distributed fusion estimator (DFE) when the S-RE channel is subject to resource constraints is given by
\begin{equation}\label{eq10}
\hat{{x}}_m(k)=\sum\limits^L_{i=1}\mathrm{W}^m_i(k)\hat{{x}}^m_i(k)
\end{equation}
where $\hat{{x}}_m(k)$ is the fusion estimate, and $\mathrm{W}^m_i(k)$ is the distributed weighting fusion matrix.

{\it{In conclusion, the case of the S-RE channel subject to resource constraints is formulated in this subsection. Then, based on the above communication strategy, this paper will address the following problems:
\begin{itemize}
  \item Firstly, for a given set of binary variables $\sigma^i_m(k)=\{\sigma_{\kappa_i}^i(k) \left|\right. \kappa_i \in \mathfrak{H}_m^i, i\in \mathfrak{L}\}$ that satisfies \eqref{re1} and \eqref{ea5}, the gain ${\mathrm{K}}^m_i(k)$ in the designed LRE (\ref{eq8}) is such that the SE of the local estimate $\hat{x}^m_i(k)$ is bounded, i.e.,
      \begin{equation}\label{am1}
        \lim_{k\rightarrow \infty} \left(x_i(k)-\hat{x}^m_i(k)\right)^{\mathrm{T}}\left(x_i(k)-\hat{x}^m_i(k)\right)< \mathcal{M}_m^i
      \end{equation}
      where $\mathcal{M}_m^i$ is a positive scalar, and then to minimize an upper bound of the SE of the corresponding LRE at each instant.
  \item Secondly, based on each LRE (\ref{eq8}), design distributed weighting fusion matrices $\{\mathrm{W}^m_i(k)\left|\right.\sum^L_{i=1}\mathrm{W}^m_i(k)=I\}$ in \eqref{eq10} such that an upper bound of the SE of the DFE is minimal at each instant.
\end{itemize}}}

{\it Remark 1}: Generally, the local estimate $\hat{{x}}^s_i(k)$ without network communication constraints can be calculated by the LNE (\ref{eq4}), while resource constraints are an inevitable problem to be considered in NMFSs. Therefore, the ETS (\ref{ea12}) and the DRS (\ref{ea5}) are introduced in a unified framework to satisfy the finite resources in the S-RE channel for the nonlinear NMFSs \eqref{eq1}-\eqref{eq2}. Notice that, the untransmitted measurements, which may cause a degradation of the estimation performance, are reconstructed by a unified compensation model \eqref{ea14}, and the compensation measurement ${z}^r_i(k)$ instead of the raw measurement ${y}_i(k)$ is used to design the estimators in this paper. In this communication strategy, the introduced ETS and DRS provide the possibility to alleviate unnecessary resource consumption, while the compensation strategy gives a desirable estimation performance of the designed estimators.

{\it Remark 2}: Since the DRS, which weights compression of measurement information to achieve dimensionality reduction, proposed in \cite{dr1,dr11,tsp1} is complicated in the calculation of the compression operator, another kind of DRS was introduced in this paper that directly selects partial components of measurements rather than the data compression approach. In addition, it can be found from \eqref{re1} that the global bandwidth constraint is assumed to be known in advance, thus a group $\sigma^i_m(k)$ can be given to satisfy \eqref{ea5} by adjusting the local constraint $\varsigma_m^i$ to satisfy \eqref{re1}, and then the dimensionality reduction matrix $\Theta^i_m(k)$ can be determined. For instance, if $m_i=4, \varsigma_m^i=2$ then $\mathfrak{h}^i_m=6$, and $\Theta^i_m(k)$ can be determined by:
\begin{equation} \label{eae2}
\begin{aligned}
&\Theta_m^i(k)=\mathrm{diag}\{\sigma^i_1(k)+\sigma^i_2(k)+\sigma^i_3(k),\\
&\qquad\qquad\qquad\;\;
\sigma^i_1(k)+\sigma^i_4(k)+\sigma^i_6(k), \\
&\qquad\qquad\qquad\;\;
\sigma^i_2(k)+\sigma^i_4(k)+\sigma^i_5(k), \\
&\qquad\qquad\qquad\;\;
\sigma^i_3(k)+\sigma^i_5(k)+\sigma^i_6(k)\}
\end{aligned}
\end{equation}
Therefore, it can be seen from \eqref{ea5} and \eqref{eae2} that $\Theta^i_m(k)$ is decided by $\sigma^i_{\kappa_i}(k)$.

{\it Remark 3}: Notice that, the innovation-based deterministic and stochastic ETSs have been well developed in \cite{b9,b17,tsp2,tsp3,tsp4,tsp5} for networked time-invariant systems, and the stochastic ETS indeed provides an important advantage that preserves the Gaussian property of the innovation sequence. Nevertheless, the stochastic ETS either requires feedback from the estimator or becomes irrelevant to the sensor model, which may increase communication consumption and be problematic in applications of time-varying systems. In fact, the considered event-based problem is in the framework of nonlinear NMFSs without a Gaussian assumption, and thus an effective deterministic ETS \eqref{ea12}-\eqref{eaa12} is proposed in this paper. Concretely, the decision variable $\gamma_m^i(k)$ is based on the difference between the raw measurement $y_i(k)$ and the latest triggered measurement $\breve{y}_i(k)$. In this case, the predefined threshold $\delta^i_m$ should be suitably selected to meet a tradeoff between the estimation performance and the resource constraints.

\subsection{Event-Triggered Scheduler Based on Local Nonlinear Estimation under Resource Constraints}
It is worth noting that smart sensors are capable of calculating local estimates in some practical applications. Then, the local estimate from each smart sensor is sent to the FC through the communication network in this kind of networked distributed fusion framework. Hence, it is essential to address the problem of the S-FC channel under restricted resources as well. Similarly, an estimate based ETS with respect to local estimation is installed in each smart sensor to meet the finite communication resources. In this scheme, the decision variable $\gamma_s^i(k) (\in \{0,1\})$ is designed based on the following event-triggered condition:
\begin{align}\label{eb9}
&\gamma_s^i(k)=\left\{
\begin{array}{lc}
0, \quad \mathrm{if} ~ \hat{x}^s_i(k) \in \mathfrak{D}_s^i(k) \\
1, \quad \mathrm{otherwise}
\end{array}
\right.  \\
&\mathfrak{D}_s^i(k)=\left\{\hat{x}^s_i(k)\left|\right.\|\hat{x}^s_i(k)-\hat{x}^l_i(k)\|_2 \leq \delta_s^i, i\in \mathfrak{L} \right\} \label{ebb9}
\end{align}
where $\delta_s^i>0$ is a predefined threshold, $\hat{x}^l_i(k)$ is the latest estimate transmitted for further dimensionality reduction processing, and $\mathfrak{D}_s^i(k)$ denotes the local estimate set of each smart sensor at untriggered moments. Specifically, $\gamma^i_s(k)=1$ indicates that the $\hat{x}^s_i(k)$ does not belong to $\mathfrak{D}_s^i(k)$ and will be processed further; otherwise, no estimate is sent to the FC at time $k$.
Therefore, the event-triggered instants sequence $0\leq t^i_s(1)\leq \cdots \leq t^i_s(k) \leq \cdots$ is determined by
\begin{equation}\label{ere12}
  t^i_s(k+1)=\min \left\{\begin{aligned} k \in \mathfrak{N}_+ \left|\right. k>t^i_s(k),\; \hat{x}^s_i(k) \notin \mathfrak{D}^i_s(k) \end{aligned}\right\}
\end{equation}

Notice that, another kind of DRS \cite{dr2} is introduced to lower estimated packet to meet the finite communication bandwidth in this subsection. By using this DRS, only partial estimates of $\gamma^i_s(k)\hat{x}^s_i(k)$ are allowed to be transmitted. Namely, only $\varsigma_s^i(\varsigma_s^i \in \mathfrak{N}_+, 1\leq \varsigma_s^i \leq n)$ components of $\hat{x}^s_i(k)$ have the opportunity to be transmitted to the FC at each instant. It is obvious that there is no estimate transmitted to the FC when $\gamma^i_s(k)=0$. Then, in terms of the mathematical description, the transmitted state estimate $\hat{x}^r_i(k)$ can only take one element from the following set:
\begin{equation}\label{eb3}
\boldsymbol{\chi}_s^i(k)\triangleq \left\{\gamma^i_s(k)\Theta^i_{\hbar_i}\hat{x}^s_i(k)\left|\right.\hbar_i\in {\mathfrak{H}}_s^i \triangleq \left\{1,2,\ldots,\mathfrak{h}_s^i\right\}\right\}
\end{equation}
where $\Theta^i_{\hbar_i}$ is also a 0-1 diagonal matrix with $\varsigma^i_s$ 1-elements, and $\mathfrak{h}_s^i=\prod^{\varsigma^i_s-1}_{\imath=0}(n-\imath)/\prod_{\jmath=1}^{\varsigma^i_s}\jmath$. Then, the transmitted estimate $\hat{x}^r_i(k)$ of each smart sensor can be formulated by
\begin{equation}\label{eb6}
\hat{x}^r_i(k)=\gamma^i_s(k)\sum\limits ^{\mathfrak{h}_s^i}_{\hbar_i=1}\sigma^i_{\hbar_i}(k)\Theta^i_{\hbar_i}\hat{x}^s_i(k)
=\gamma^i_s(k)\Theta_s^i(k)\hat{x}^s_i(k)
\end{equation}
where $\sigma^i_{\hbar_i}(k)(\in \{0,1\})$ is a binary variable describing the dimensionality status, and the   $\sum_{\hbar_i=1}^{\mathfrak{h}_s^i}\sigma^i_{\hbar_i}(k)=1$ should be held. Then, ${\Theta}_s^i(k)$ is one of elements in the set $\{{\Theta}^i_{\hbar_i}\left|\right. \hbar_i \in \mathfrak{H}^i_s\}$, and depend on the choice of corresponding $\sigma^i_{\hbar_i}(k)$.

Moreover, it can be found from the above DRS that the practical communication status is related to the variable $\sigma^i_{\hbar_i}(k)$ at each instant. In this sense, the stochastic process $\{\sigma^i_{\hbar_i}(k)\}$ is assumed to be independent and identically distributed (i.i.d.) \cite{qd1}, i.e.,
\begin{equation}
\begin{aligned}\label{eb17}
&\mathbb{E}\{\sigma^i_{\hbar_i}(k)\sigma^j_{\hbar^0_j}(k_0)\} \\
&=\left\{
\begin{array}{ll}
0,  &i=j,k=k_0,\hbar_i \neq \hbar^0_i \\
\mathbb{E}\{\sigma^i_{\hbar_i}(k)\},  &i=j,k=k_0, \hbar_i = \hbar^0_i \\
\mathbb{E}\{\sigma^i_{\hbar_i}(k)\}\mathbb{E}\{\sigma^j_{\hbar^0_j}(k_0)\}, &\forall \;  i,j,k,k_0,\hbar_i,\hbar^0_i
\end{array}
\right.
\end{aligned}
\end{equation}
where $\mathrm{Pr}\{\sigma^i_{\hbar_i}(k)=1\}=\pi^i_{\hbar_i}$ is the known selected probability and satisfying
\begin{equation}
\label{eb15}
 \sum\nolimits^{\mathfrak{h}_s^i}_{\hbar_i=1}\pi^i_{\hbar_i}= 1
\end{equation}
It should be noted that the local bandwidth constraint $\varsigma^i_s$ is known instead of global bandwidth constraints in this communication strategy, then a group of $\sigma^i_s(k)=\{\sigma_{\hbar_i}^i(k) \left|\right. \hbar_i \in \mathfrak{H}_s^i, i\in \mathfrak{L}\}$ will be generated based on the selection probabilities $\pi^i_{\hbar_i}(\hbar_i\in \mathfrak{H}_s^i) $, which gives a different way for determining dimensionality matrix $\Theta^i_s(k)$.

\begin{figure}[!htb]
\begin{center}
  \setlength{\abovecaptionskip}{0.1cm}
  \centering
  \includegraphics[width=9cm, height=5cm]{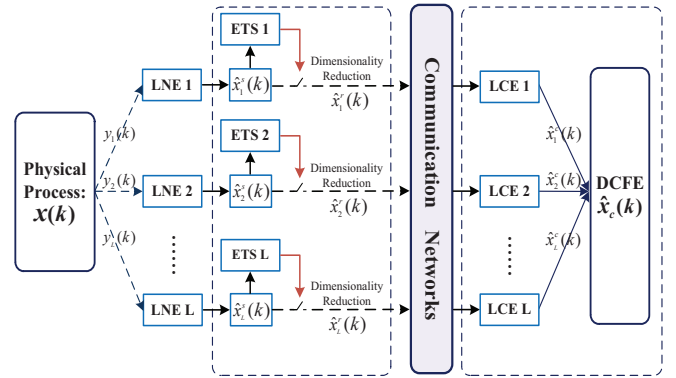}
\caption{The diagram of modeling process when the S-FC channel is subject to resource constraints.}
\label{fig2}
  \end{center}
\end{figure}

Similarly, to improve the accuracy of the fusion estimation, a unified compensation model with respect to local estimation is developed to address this problem. Then, the local compensation state estimator (CSE) is modeled by
\begin{equation}\label{eb12}
\hat{x}^c_i(k)=\hat{x}^r_i(k)+(I-\gamma_s^i(k)\Theta_s^i(k))f(\hat{x}^c_i(k-1))
\end{equation}
where $\hat{x}^c_i(k)$ is the compensatory estimate, $f(\hat{x}_i^c(k-1))$ is the one-step prediction of the local compensation estimation. When $\gamma_s^i(k)=1$, it can be seen from \eqref{eb12} that the dimensionality reduction estimate $\hat{x}^r_i(k)$ will be transmitted to the FC, and $(I-\Theta_s^i(k))f(\hat{x}_i^c(k-1))$ is utilized to compensate the untransmitted components of $\hat{x}^r_i(k)$; otherwise, the prediction $f(\hat{x}_i^c(k-1))$ is employed to compensate the untransmitted local estimate directly. The specific distributed compensation fusion estimation process based on the above communication
strategy can be shown in Fig. \ref{fig2}.

Next, when the S-FC channel is subject to resource constraints, the distributed compensation fusion estimator (DCFE) based on the local CSE \eqref{eb12} is given by
\begin{equation}\label{eb13}
\hat{x}_c(k)=\sum^L_{i=1}\mathrm{W}^s_i(k)\hat{x}^c_i(k)
\end{equation}
where $\hat{x}_c(k)$ is the compensatory fusion estimate, and $\mathrm{W}^s_i(k)$ is the weighting fusion matrix to be determined.

{\it{In conclusion, the case of the S-FC channel subject to resource constraints is considered in this subsection. For a different communication strategy mentioned above, the following problems in this paper will be addressed:
\begin{itemize}
  \item Firstly, for each local CSE (\ref{eb12}), the selection probabilities $\{\pi^i_{\hbar_i}\left|\right.\sum^{\mathfrak{h}_s^i}_{\hbar_i=1}\pi^i_{\hbar_i}=1\}$ in \eqref{eb15} should be determined such that the mean square error (MSE) of the CSE is bounded, i.e.
      \begin{equation}\label{bm1}
        \lim_{k\rightarrow \infty}\mathbb{E}\left(\left(x(k)-x^c_i(k)\right)^{\mathrm{T}}\left(x(k)-x^c_i(k)\right)\right)< \mathcal{M}_s^i
      \end{equation}
      where $\mathcal{M}_s^i$ is a positive scalar.
  \item Secondly, the distributed weighting fusion matrices $\{\mathrm{W}^s_i(k)\left|\right.\sum^L_{i=1}\mathrm{W}^s_i(k)=I\}$ in \eqref{eb13} should be determined such that an upper bound of the MSE of the DCFE is minimal at each instant.
\end{itemize}}}

{\it Remark 4}: When considering that the S-FC channel is subject to resource constraints, an unfixed structure DRS \cite{dr2} was introduced in this section. Since the ${\Theta}_s^i(k)$ in \eqref{eb6} is decided by the $\sigma^i_{\hbar_i}(k)$, and the $\{\sigma^i_{\hbar_i}(k)\}$ is an i.i.d. stochastic process, thus ${\Theta}_s^i(k)$ is a random matrix. In this case, \cite{dr2} proposed a distributed fusion Kalman filter for linear NMFSs under communication constraints, while the system noise covariances need to be known a priori. To overcome this limitation, a bounded recursive optimization scheme has been developed in \cite{qd1} to deal with the bounded noises, in which a QS and the same DRS were utilized to alleviate the communication burden. However, the design of event-triggered estimators for distributed nonlinear NMFSs with unknown noise statistical properties is still a challenging issue. Therefore, the ETS \eqref{eb9}-\eqref{ebb9} and the DRS \eqref{eb3}-\eqref{eb15} are proposed with a unified framework for the nonlinear NMFSs without noise statistical properties, and then the nonlinear compensation estimators \eqref{eb12} and \eqref{eb13} are designed to ensure a satisfactory estimation performance.

{\it Remark 5}: Though the DRS and the ETS were used in \cite{b18} to address the networked fusion estimation problems under resource constraints, it was still in a linear framework with the Gaussian noise assumption. Moreover, the local estimate was utilized to compensate for the untransmitted estimate in the FC, while it seems unrealistic to have a complete local estimate in the FC under this communication strategy. Indeed, the one-step prediction $f(\hat{x}^c_i(k-1))$ of the local compensation estimate $\hat{x}^c_i(k-1)$ is ideal as the compensation information, which can be obtained in the FC. Moreover, the ETS in \cite{b18} was based on the dimensionality reduction estimation, while the ETS in this paper is based on the original local estimation directly. Under the proposed communication strategy, the DRS \eqref{eb3}-\eqref{eb6} is not required to work when $\gamma^i_s(k)=0$ (i.e., the event is not triggered at time $k$), which can save unnecessary resource consumption.

\section{Main Results}
Before deriving the main results, a useful Lemma is introduced as follows:

{\it Lemma 1}: \label{le1} \cite{b35} Let $S^{\mathrm{T}}_1=S_1$, $S_2$ and $S_3$ be real matrices of appropriate dimensions with $P(k)$ satisfying $P^{\mathrm{T}}(k)P(k) \leq I$. Then
$$S_1+S_3P(k)S_2+S^{\mathrm{T}}_2P^{\mathrm{T}}(k)S^{\mathrm{T}}_3<0$$
if and only if there exists a positive scalar $\varepsilon >0$ such that
$$
\left[
\begin{array}{ccc}
-\varepsilon I & \varepsilon S_2 & 0 \\
{*} & S_1 & S_3 \\
{*} & {*} & -\varepsilon I
\end{array}
\right]<0
$$

\subsection{Distributed Fusion Estimation Based on Compensation Measurement for Nonlinear NMFSs}
When considering the estimation problems of the S-RE channel under resource constraints, a deterministic ETS and a directly DRS are introduced in Sec.II-A to meet the finite communication resources. Then, according to the designed LRE \eqref{eq8} and DFE \eqref{eq10}, the problems of determining the local estimator gain ${\mathrm{K}}^m_i(k)$ and the distributed weighting fusion matrix $\mathrm{W}^m_i(k)$ for nonlinear systems (\ref{eq1}) and (\ref{ea14}) are solved in Theorem 1.

Let $\tilde{x}^{-}_{m_i}(k) \triangleq {x}(k)-{\hat{x}}^{-}_{m_i}(k)$ and $\tilde{x}^m_i(k) \triangleq {x}(k)-{\hat{x}}^m_i(k)$ denote the prediction error and local estimation error, respectively. Then, it follows from (\ref{eq1}) and (\ref{eq8}) that
\begin{equation}\label{eq12}
\left \{
\begin{aligned}
&\tilde{x}^{-}_{m_i}(k)={f}({x}(k-1))\\
&\qquad \qquad -{f}_i(\hat{x}^m_i(k-1))+{\Gamma}(k-1){w}(k-1)\\
&\tilde{x}^m_i(k)=\tilde{x}^{-}_{m_i}(k)-{\mathrm{K}}^m_i(k)\left[z^r_i(k)-h_i(\hat{x}^-_{m_i}(k))\right]
\end{aligned}
\right.
\end{equation}
To further analyze the above nonlinear estimation error, the nonlinear functions ${f}(x(k-1))$ and ${h_i}(x(k))$ in \eqref{eq12} are linearized by using the first order Taylor series expansion (TSE), and then the higher-order terms are modeled by state-dependent matrices with uncertain parameters \cite{b36}, one has
\begin{equation}\label{eq13}
\left\{
\begin{aligned}
&{f}(x(k-1))={f}_i(\hat{x}^m_i(k-1)) \\
&\qquad \qquad \qquad +({A}_{f_i}^m(k-1)+{M}_{f}^i(k){N}_{f}^i(k))\tilde{x}^m_i(k-1)  \\
&{h}_i(x(k))={h}_i(\hat{x}^-_{m_i}(k))+({C}_{h_i}^m(k)+{M}_{h}^i(k){N}_{h}^i(k))\tilde{x}^{-}_{m_i}(k) \\
\end{aligned}
\right.
\end{equation}
where ${M}_{f}^i(k)>0$ and ${M}_{h}^i(k)>0$ are state-dependent scaling matrices, ${N}_{f}^i(k)$ and ${N}_{h}^i(k)$ are unknown bounded matrices that satisfy $\|{N}_{f}^i(k)\|_2\leq 1$ and $\|{N}_{h}^i(k)\|_2\leq 1$, respectively. Moreover, ${A}_{f_i}^m(k-1)$ and ${C}_{h_i}^m(k)$ are Jacobin matrices in first-order TSE of ${f}(x(k-1))$ and ${h_i}(x(k))$, respectively, which are defined in Table \uppercase\expandafter{\romannumeral2} of Appendix C.

Then, substituting (\ref{eq13}) into (\ref{eq12}), the estimation error $\tilde{x}^m_i(k)$ is written as:
\begin{equation}\label{eq16}
\begin{aligned}
\tilde{x}^m_i(k)&=(\mathrm{K}_{C_i}^{m}(k){A}_{f_i}^m(k-1)+\mathrm{K}^{m}_{C_i}(k){M}_{f}^i(k){N}_{f}^i(k)\\
&-\mathrm{K}^m_{\Theta_i}(k){M}_{h}^i(k){N}_{h}^i(k){A}_{f_i}^m(k-1)-\alpha^i_m(k) \\
&\times \mathrm{K}^m_{\Theta_i}(k){M}_{h}^i(k){N}_m^i(k))\tilde{x}^m_i(k-1) +(\mathrm{K}_{C_i}^{m}(k)\\
&-\mathrm{K}^m_{\Theta_i}(k){M}_{h}^i(k){N}_{h}^i(k)){\Gamma}(k-1){w}(k-1)\\
&-\gamma_m^i(k)\mathrm{K}^m_i(k)\Theta_m^i(k){D}_i(k){v}_i(k)
\end{aligned}
\end{equation}
where $\mathrm{K}_{C_i}^{m}(k)$, $\mathrm{K}^m_{\Theta_i}(k)$ and $N_m^i(k)$ are defined in TABLE \uppercase\expandafter{\romannumeral2}, $\alpha^i_m(k)$ is a scalar that is equal or greater than the max element of ${M}_{f}^i(k)$. 

{\it Theorem 1}:  \label{the1} For a predefined threshold $\delta_m^i$ satisfying \eqref{ea12}-\eqref{eaa12}, and a given global bandwidth constraint $\varsigma_m$ and a set of binary variables $\sigma^i_m(k)$ satisfying \eqref{re1}-\eqref{ea5}, the following convex optimization problem is being developed to determine each LRE gain $\mathrm{K}^m_i(k)$ for nonlinear systems (\ref{eq1}) and (\ref{ea14}):
\begin{equation}\label{eq18}
\begin{array}{l}
\mathop {\min}\limits_{\Psi_i(k)>0,\Upsilon_i(k)>0,\Phi_i(k)>0 \hfill \atop
 \mathrm{K}^m_i(k),\epsilon_{1i}(k),\epsilon_{2i}(k),\epsilon_{3i}(k)\hfill}\mathrm{Tr}\{\Phi_i(k)\}+ \mathrm{Tr}\{\Upsilon_i(k)\} \\
\mathrm{s.t.}:\left\{
\begin{array}{lllll}
\left[
  \begin{array}{ccc}
    -\epsilon_{1i}(k)I \!&\! O_I^i(k) \!&\! 0 \\
     * \!&\! -\Delta_m^i(k) \!&\! \mathrm{K}^M_{C_i}(k)\\
      * \!&\! * \!&\! -\epsilon_{1i}(k)I
  \end{array}
\right]<0 \\
\left[
  \begin{array}{ccc}
    -\epsilon_{2i}(k)I \!&\! O_A^i(k) \!&\! 0 \\
     * \!&\! -\Delta_m^i(k) \!&\! \mathrm{K}_{\Theta_i}^M(k)\\
      * \!&\! * \!&\! -\epsilon_{2i}(k)I
  \end{array}
\right]<0 \\
\left[
  \begin{array}{ccc}
    -\epsilon_{3i}(k)I \!&\! O_{\alpha}^i(k) \!&\! 0 \\
     * \!&\! -\Delta_m^i(k) \!&\! \mathrm{K}_{\Theta_i}^i(k)\\
      * \!&\! * \!&\! -\epsilon_{3i}(k)I
  \end{array}
\right]<0 \\
\Psi_i(k)-\zeta_i(k)I<I \\
\zeta_i(k)< 1
\end{array}
\right.\\
\end{array}
\end{equation}
where $O_I^i(k)$, $O_A^i(k)$, $O_{\alpha}^i(k)$, $\mathrm{K}^M_{C_i}(k)$ and $\mathrm{K}_{\Theta_i}^M(k)$ are defined in TABLE II. Then, the SE of the local estimate $\hat{x}^m_i(k)$ in (\ref{eq8}) will be bounded, i.e.,
\begin{equation}\label{ee19}
\lim\limits_{k\rightarrow \infty}\left(\tilde{{x}}^m_i(k)\right)^{\mathrm{T}}\tilde{{x}}^m_i(k)<\mathcal{M}_m^i
\end{equation}
where $\mathcal{M}_m^i$ is a positive scalar. Moreover, a convex optimization problem is being established to obtain
distributed weighting fusion matrix $\mathrm{W}^m_i(k)$ as follows:
\begin{equation}\label{eq30}
\begin{array}{l}
\mathop{\min}\limits_{\Psi(k)>0,\Phi(k)>0,\Upsilon(k)>0\hfill \atop
\mathrm{W}_m(k),\Psi_1(k),\Psi_2(k),\Phi_1(k) \hfill}\mathrm{Tr} \{\Psi(k)\}+\mathrm{Tr} \{\Phi(k)\}+\mathrm{Tr} \{\Upsilon(k)\} \\
\mathrm{s.t.}:\left\{
\begin{array}{lllll}
\left[
  \begin{array}{ccc}
    -\epsilon_{1}(k)I \!&\! O^m_I(k) \!&\! 0 \\
     * \!&\! -\Delta_m(k) \!&\! \mathrm{K}^{\mathrm{W}}_F(k)\\
      * \!&\! * \!&\! -\epsilon_{1}(k)I
  \end{array}
\right]<0 \\
\left[
  \begin{array}{ccc}
    -\epsilon_{2}(k)I \!&\! O^m_A(k) \!&\! 0 \\
     * \!&\! -\Delta_m(k) \!&\! \mathrm{K}^{\mathrm{W}}_H(k)\\
      * \!&\! * \!&\! -\epsilon_{2}(k)I
  \end{array}
\right]<0 \\
\left[
  \begin{array}{ccc}
    -\epsilon_{3}(k)I \!&\! O^m_{\alpha}(k) \!&\! 0 \\
     * \!&\! -\Delta_m(k) \!&\! \mathrm{K}^{\mathrm{W}}_H(k)\\
      * \!&\! * \!&\! -\epsilon_{3}(k)I
  \end{array}
\right]<0\\
\end{array}
\right.\\
\end{array}
\end{equation}
where $O^m_I(k)$, $O^m_A(k)$, $O^m_{\alpha}(k)$, $\mathrm{K}^{\mathrm{W}}_F(k)$ and $\mathrm{K}^{\mathrm{W}}_H(k)$ are defined in TABLE II.

{\it Proof} : See the proof in Appendix A.

Based on the analysis of the state estimation problem of the NMFSs under the resource constraints of the S-E communication channel in this subsection, the local estimate ${\hat{{x}}^m_{i}}(k)$ in (\ref{eq8}) and the fusion estimate ${\hat{x}_m(k)}$ in (\ref{eq10}) can be obtained by implementing the Algorithm 1.
\begin{table}[htbp]
  \normalsize
	\centering
	\label{tab1}
	\begin{tabular}{p{0.92\columnwidth}}
		\toprule
		\textbf{Algorithm 1}: Distributed Networked Fusion Estimation under S-RE Channel Resource Constraints \\[-0.25mm]
		\midrule \vspace{-8pt}
		\begin{itemize}
        \item[1:] Initialization states ${x}(0)$ and $\hat{x}_i^m(0)(i \in \mathfrak{L})$;
        \item[2:] \textbf{for} $i:=1$ \textbf{to} $L$ \textbf{do}
        \item[3:] Given a predefined threshold $\delta_m^i$ to calculate $\gamma^i_m(k)$, and choose a group of binary variables $\sigma^i_m(k)$ to determine $\Theta^i_m(k)$;
		\item[4:] Solve the optimization problem (\ref{eq18}) by using the $``mincx"$ function of MATLAB LMI Toolbox,
                  then determine LRE gain $\mathrm{K}^m_i(k)$;
        \item[5:] Calculate ${\hat{x}^m_{i}}(k)$ by (\ref{eq8});
        \item[6:] \textbf{end for}
		\item[7:] Based on each ${\hat{x}^m_{i}}(k)$, determine weighting fusion matrices ${\mathrm{W}^m _i}(k)(i\in \mathfrak{L})$ by solving (\ref{eq30});
		\item[8:] Calculate fusion estimate ${\hat{x}_m(k)}$ by (\ref{eq10});
		\item[9:] Return to step 2 and repeat steps 2-8 to calculate ${\hat {x}}^m_i(k+1)$ and ${\hat{x}_m(k+1)}$.
		\end{itemize}\\ [-4mm]
		\bottomrule
	\end{tabular}
\end{table}

{\it Remark 6}: For nonlinear cyber-physical systems with bounded noises, a security fusion estimation problem subject to DoS attacks has been investigated in \cite{q1}. Since the linearization errors caused by the TSE were modeled as bounded noises, then the stability of the designed estimator has not been addressed in \cite{q1}. Indeed, the neglected linearization errors can have an influence on the stability analysis and the estimation precision of the estimator. Notice that, this paper introduces the state-dependent matrices and the unknown bounded parameters to model the high-order terms of the TSE in \eqref{eq13}. In this case, the stability of the designed nonlinear estimator \eqref{eq12} can be further analyzed, and then the stability condition such that the SE of the local/fusion estimators is bounded is presented in Theorem 1.

\subsection{Distributed Fusion Estimation Based on Local Compensation Estimate for Nonlinear NMFSs}
Since the networked fusion estimation problem of S-FC channel under resource constraints is also taken into account, and  then an kind of DRS and ETS are introduced in Sec.II-B. In this subsection, the process for determining the distributed weighting fusion matrix $\mathrm{W}^s_i(k)$ based on CSE \eqref{eb12} and DCFE \eqref{eb13} is presented in Theorem 2.

Notice that, each local estimate $\hat{x}^s_i(k)$ can be calculated by using \cite[Th.1]{q2} based on LNE \eqref{eq4}.
Then, the local estimation error $\tilde{x}^s_i(k)=x(k)-\hat{x}^s_i(k)$ is given by
\begin{equation}\begin{aligned}\label{ec1}
\tilde{x}^s_i(k)=&\;{A}_s^i(k)\tilde{x}^s_i(k-1)+{\Gamma}_s^i(k)w(k-1)\\
&-\mathrm{K}^s_i(k){D}_i(k)v_i(k) \\
\end{aligned}\end{equation}
Meanwhile, let $\tilde{x}^c_i(k)=x(k)-\hat{x}^c_i(k)$ denote CSE error, one has from
(\ref{eq1}), (\ref{eb12}) and \eqref{ec1} that
\begin{equation}\begin{aligned}\label{ec3}
&\tilde{x}^c_i(k)={A}_{\theta}^i(k)\tilde{x}_i^c(k-1)+\gamma^i_s(k)\Theta_s^i(k)A_s^i(k)\tilde{x}^s_i(k-1)\\
&\;\;\;\;+{\Gamma}_c^i(k)w(k-1)-\gamma_s^i(k)\Theta_s^i(k)\mathrm{K}^s_i(k)D_i(k)v_i(k)\\
\end{aligned}\end{equation}
where ${A}_s^i(k), {\Gamma}_s^i(k), {A}_{\theta}^i(k)$, and ${\Gamma}_c^i(k)$ are defined in TABLE III of Appendix C.

Similarly, the nonlinear functions in the derivation process of \eqref{ec1} and \eqref{ec3} are also addressed by first order TSE, in which the linearization errors are modeled by state-dependent matrices with uncertain matrices. Then, the introduced matrices ${L}_{{f}}^i(k)$, ${L}_{h}^i(k)$, and ${L}^i_{c}(k)$ are defined as positive state-dependent matrices, the unknown matrices ${P}_{f}^i(k)$, ${P}_{h}^i(k)$, and ${P}^i_{c}(k)$ are assumed to satisfy $\|{P}_{f}^i(k)\|_2\leq 1$, $\|{P}_{h}^i(k)\|_2\leq 1$, and $\|{P}^i_{c}(k)\|_2\leq 1$, respectively. Meanwhile, $\alpha^i_s(k)$ is a scalar that is equal or greater than the max element of ${L}_{f}^i(k)$. 

{\it Theorem 2}: For the given triggering threshold $\delta_s^i>0$, if there exist integer $N_i\geq 0$ and $\rho_i(k)>0$ such that the selection probability $\pi^i_{\hbar_i}$ in (\ref{eb15}) satisfying
\begin{equation}\label{ec18}
\begin{bmatrix} -\rho_i(k)I & 0 & \rho_i(k)I & 0 \\
* & -I & \Theta_I^iA^i_c(k-1) & \Theta_I^iL^i_c(k)\\
* & * & -I & 0 \\
* & * & * & -\rho_i(k)I
\end{bmatrix} <0
\end{equation}
\begin{equation}\begin{aligned}\label{ec28}
\|\breve{\mathcal{R}_i}(k-N_i,\breve{\mathcal{R}_i}(k-N_i-1,\breve{\mathcal{R}_i}(\ldots,\breve{\mathcal{R}_i}(k,I))))\|_2<1
\end{aligned}\end{equation}
Then, the MSE of the CSE \eqref{eb12} is bounded. Moreover, an convex optimization problem is being constructed to calculate distributed weighting fusion matrix $\mathrm{W}^m_i(k)$ as follows:
\begin{equation}\label{ecc1}
\begin{array}{l}
\mathop{\min}\limits_{\Lambda(k)>0,\Xi(k)>0,\Sigma(k)>0, \hfill\atop
{\Xi}_1(k),{\Xi}_2(k),{\Lambda}_1(k), \mathrm{W}_s(k)\hfill}\mathrm{Tr} \{{\Xi}(k)\}+\mathrm{Tr} \{{\Lambda}(k)\}+\mathrm{Tr} \{{\Sigma}(k)\} \\
\mathrm{s.t.}:\left\{
\begin{array}{llllll}
\left[
  \begin{array}{ccc}
    -\varrho_{1}(k)I \!&\! {O}^s_I(k) \!&\! 0 \\
     * \!&\! -{\Delta}_c(k) \!&\! {L}^{\mathrm{W}}_{\mathrm{K}}(k)\\
      * \!&\! * \!&\! -\varrho_{1}(k)I
  \end{array}
\right]<0 \\
\left[
  \begin{array}{ccc}
    -\varrho_{2}(k)I \!&\! {O}^s_A(k) \!&\! 0 \\
     * \!&\! -{\Delta}_c(k) \!&\! \mathrm{K}^{\mathrm{W}}_{\Theta}(k)\\
      * \!&\! * \!&\! -\varrho_{2}(k)I
  \end{array}
\right]<0 \\
\left[
  \begin{array}{ccc}
    -\varrho_{3}(k)I \!&\! {O}^s_L(k) \!&\! 0 \\
     * \!&\! -{\Delta}_c(k) \!&\! \mathrm{K}^{\mathrm{W}}_{\Theta}(k)\\
      * \!&\! * \!&\! -\varrho_{3}(k)I
  \end{array}
\right]<0\\
\left[
  \begin{array}{ccc}
    -\varrho_{4}(k)I \!&\! {O}^s_\mathrm{W}(k) \!&\! 0 \\
     * \!&\! -{\Delta}_c(k) \!&\! {L}^{\mathrm{W}}_{\Theta}(k)\\
      * \!&\! * \!&\! -\varrho_{4}(k)I
  \end{array}
\right]<0\\
\end{array}
\right.
\end{array}
\end{equation}
where ${O}^s_I(k), {O}^s_A(k), {O}^s_L(k), {O}^s_\mathrm{W}(k), {L}^{\mathrm{W}}_{\mathrm{K}}(k), \mathrm{K}^{\mathrm{W}}_{\Theta}(k), {L}^{\mathrm{W}}_{\Theta}(k)$ and ${\Delta}_c(k)$ are defined in TABLE III.

{\it Proof}: See the proof in Appendix B.

Based on the above analysis of the fusion estimation problem of the NMFSs under the case of the S-FC channel being subject to resource constraints in this subsection, the compensatory fusion estimate $\hat{x}_c(k)$ can be calculated by implementing the Algorithm 2.
\begin{table}[htbp]
\normalsize
	\centering
	\begin{tabular}{p{0.92\columnwidth}}
		\toprule
		\textbf{Algorithm 2}:  Distributed Compensation Fusion Estimation under S-FC Channel Resource Constraints \\[0.25mm]
		\midrule \vspace{-8pt}
		\begin{itemize}
		\item[1:] Initialization states ${x}(0)$ and $\hat{x}_i^c(0)$;
        \item[2:] \textbf{for} $i:=1$ \textbf{to} $L$ \textbf{do}
        \item[3:] Calculate LNE gain $\mathrm{K}^s_i(k)$ and each local estimates ${\hat{{x}}^s_{i}}(k)$ by \cite[Th.1]{q2};
        \item[4:] Given a predefined threshold $\delta_s^i$ to calculate $\gamma^i_s(k)$, and select probabilities $\pi^i_{\hbar_i}(\hbar_i\in \mathfrak{H}_s^i)$ to randomly generate a group of binary variables $\sigma^i_s(k)$ to determine $\Theta^i_s(k)$;
		\item[5:] Calculate local compensatory estimate ${\hat{{x}}^c_{i}}(k)$ by (\ref{eb12});
        \item[6:] Solve the optimization problem (\ref{ecc1}) to determine weighting fusion matrices ${\mathrm{W}^s _i}(k)(i\in \mathfrak{L})$;
		\item[7:] Calculate fusion estimate ${\hat{x}_c(k)}$ by (\ref{eb13});
        \item[8:] \textbf{end for}
		\item[9:] Return to step 2 and repeat steps 2-8 to calculate ${\hat{x}_c(k+1)}$.
		\end{itemize}\\ [-4mm]
		\bottomrule
	\end{tabular}
\end{table}

{\it Remark 7}: Notice that, each smart sensor can generate a group of i.i.d. stochastic variables $\sigma_{\hbar_i}^i(k)(\hbar_i\in \mathfrak{H}_s^i)$ for the given selection probabilities $\pi^i_{\hbar_i}(\hbar_i\in \mathfrak{H}_s^i)$, which is based on \eqref{eb17}-\eqref{eb15}. Therefore, when $\pi^i_{\hbar_i}$ is selected from the developed stability conditions (\ref{ec18})-(\ref{ec28}), the information transmission matrix ${\Theta}_s^i(k)$ in \eqref{eb6} can be determined because they are decided by $\sigma_{\hbar_i}^i(k)$. Then, the CSE \eqref{eb12} can be calculated based on the decision variable $\gamma^i_s(k)$ and dimensionality matrix $\Theta^i_s(k)$. In this case, the probability-dependent selection criterion provides a pledge such that the MSEs of the designed CSE \eqref{eb12} and DCFE \eqref{eb13} are bounded based on the proposed ETS \eqref{eb9}-\eqref{ebb9} and DRS \eqref{eb3}-\eqref{eb15} in this paper.

\section{Simulation example}
\renewcommand\arraystretch{1.5}
Consider a vehicle localization system in the 2D horizonal space, and the vehicle's motion state is defined by ${x}(k) \triangleq {\mathrm{col}}\{p_x(k), p_y(k),\theta(k)\}$, where $p_x(k)$ and $p_y(k)$ are the position of vehicle along X and Y axis, respectively, and $\theta(k)$ denotes the heading angle. Then, the motion model of the moving vehicle can be given by \cite{q1}:
\begin{equation} \label{s1}
\left\{
\begin{aligned}
&p_x(k+1)=p_x(k)+\frac{\breve{c}_t(k)}{\breve{c}_r(k)}\cos\left(\theta(k)+\frac{t_0\breve{c}_r(k)}{2}\right)\\
&p_y(k+1)=p_y(k)+\frac{\breve{c}_t(k)}{\breve{c}_r(k)}\sin\left(\theta(k)+\frac{t_0\breve{c}_r(k)}{2}\right)  \\
&\theta (k+1)=\theta(k)+t_0\breve{c}_r(k) \\
&\breve{c}_t(k)=c_t+w_t(k)\\
&\breve{c}_r(k)=c_r+w_r(k)
\end{aligned}\right.
\end{equation}
where $c_t$ and $c_r$ are the motion commands to control the translational velocity and rotational velocity, respectively. ${w}_t(k)$ and ${w}_r(k)$ are bounded disturbances, $t_0$ is the sampling period.

In addition, six distance sensors divided into two groups are used to track the moving vehicle, then the measurement information can be obtained by:
\begin{equation}\label{s2}
	y_i(k)=\begin{bmatrix}
  \sqrt{(p_{x}(k)-{p^i_{x_1}})^2+(p_{y}(k)-{p^i_{y_1}})^2} \\
  \sqrt{(p_{x}(k)-{p^i_{x_2}})^2+(p_{y}(k)-{p^i_{y_2}})^2} \\
  \sqrt{(p_{x}(k)-{p^i_{x_3}})^2+(p_{y}(k)-{p^i_{y_3}})^2}
	       \end{bmatrix}+D_iv_i(k)
\end{equation}
where $(p^i_{x_j},{p^i_{y_j}})(i=1,2;\;j=1,2,3)$ are the positions of these sensors in the X-Y plane.
Then, it is easily deduced from \cite{q1} that the localization system \eqref{s1}-\eqref{s2} can be formulated as (\ref{eq1})-(\ref{eq2}), and the detailed conversion process is omitted here.

In this example, some common parameters are set as $t_0=1.8$, $c_t=0.7$, $c_r=0.8$; the measurement noise matrices $D_1=\mathrm{diag}\{0.7,0.6,0.5\}$, $D_2=\mathrm{diag}\{0.6,0.7,0.8\}$, and the system noises ${w}_t(k)$, ${w}_r(k)$ and ${v}_i(k)(i=1,2)$ are given by:
\begin{equation} \label{s3}
\left\{\begin{aligned}
&{w}_t(k)=0.3\beta_{t}(k)-0.1; {w}_r(k)=0.2 \beta_{r}(k)-0.1;\\
&{v}_{1}(k)\;={\mathrm{col}} \left\{0.3 \beta_{11}(k)-0.2, 0.2 \beta_{{12}}(k)-0.1,\right.\\
&\quad\quad \quad \quad \quad \;\;\left.0.4 \beta_{{13}}(k)-0.1\right\}; \\
&{v}_2(k)\;={\mathrm{col}}\left\{0.2 \beta_{{21}}(k)-0.1, 0.5 \beta_{{22}}(k)-0.3,\right.\\
&\quad\quad \quad \quad \quad \;\;\left. 0.4 \beta_{{23}}(k)-0.2\right\}.
\end{aligned}\right.
\end{equation}
where $\beta_{t}(k)$, $\beta_{r}(k)$, and $\beta_{ij}(k)(i=1,2;\;j=1,2,3)$ are uniformly distributed random variables over [0,1]. Meanwhile, the coordinates of six distance sensors are installed at $({p^1_{x_1}},{p^1_{y_1}})=(-25,-5)$, $({p^1_{x_2}},{p^1_{y_2}})=(-30,15)$, $({p^1_{x_3}},{p^1_{y_3}})=(-25,35)$, $({p^2_{x_1}},{p^2_{y_1}})=(25,-5)$, $({p^2_{x_2}},{p^2_{y_2}})=(30,15)$, $({p^2_{x_3}},{p^2_{y_3}})=(25,35)$, respectively.

\subsection{ETS and DRS Based on Measurement Information}
When the S-RE channel is suffering from resource constraints, the compensation measurement \eqref{ea14} was used to estimate the system state \eqref{s1} by the theoretical analysis in Sec.II-A. Based on the ETS \eqref{ea12}-\eqref{eaa12}, the triggering thresholds in case A are set to $\delta^1_m=\delta^2_m=1.2$, respectively. Since there are three communication channels in the measurement model \eqref{s2}, assuming the global bandwidth channel is $\varsigma_m=4$ in this example, and then the local bandwidth $\varsigma^i_m(i=1,2)$ can be chosen from the following set to satisfy \eqref{re1}:
\begin{equation}\label{SA1}
  \mathcal{Q}_m=\left\{\left(\varsigma^1_m,\varsigma^2_m\right)\left|\right.(2,2),(1,3),(3,1)\right\}
\end{equation}
If the bandwidth group $(2,2)$ is chosen, each measurement group has $\mathfrak{h}^1_m=\mathfrak{h}^2_m=3$ different kinds of transmission statuses. In particular, it can be concluded from Remark 2 that the dimensionality reduction matrix $\Theta_m^i(k)$ is determined by $\sigma^i_m(k)$, i.e.:
\begin{equation} \label{SA2}
\left\{\begin{aligned}
&\Theta_m^1(k)=\mathrm{diag}\{\sigma^1_1(k)+\sigma^1_2(k),\sigma^1_1(k)\\
&\qquad\qquad\quad\;
+\sigma^1_3(k),\sigma^1_2(k)+\sigma^1_3(k)\}; \\
&\Theta_m^2(k)=\mathrm{diag}\{\sigma^2_1(k)+\sigma^2_2(k),\sigma^2_1(k)\\
&\qquad\qquad\quad\;
+\sigma^2_3(k),\sigma^2_2(k)+\sigma^2_3(k)\}. \\
\end{aligned}\right.
\end{equation}
Thus, the communication matrix $\Theta_m^i(k)$ can be determined at each time. Similarly, by selecting the local bandwidth constraints from the $\mathcal{Q}_m$, the corresponding $\Theta_m^i(k)$ can be calculated. Moreover, the introduced matrices in \eqref{eq16} are given by ${M}^1_{f}=\mathrm{diag}\{0.02,0.01,0.03\}$, ${M}^2_{f}=\mathrm{diag}\{0.03,0.01,0.02\}$, ${M}_{h}^1=\mathrm{diag}\{0.03,0.03,0.02\}$, ${M}_{h}^2=\mathrm{diag}\{0.03,0.02,0.02\}$,
 and $\alpha^1_m=\alpha^2_m=1$.

\begin{figure}[!htb]
	\setlength{\abovecaptionskip}{0cm}
	\centering
	\includegraphics[width=9cm,height=7cm]{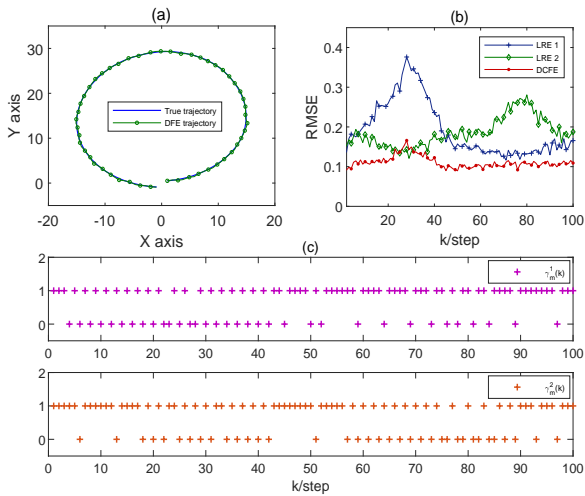}
	\caption{(a) The vehicle's true motion trajectory and fusion estimation trajectory; (b) The estimation performance comparison between the LREs and DFE; (c) The triggering status of two sensor groups.}
	\label{fig3}
\end{figure}
\begin{figure}[!htb]
	\setlength{\abovecaptionskip}{0cm}
	\centering
	\includegraphics[width=9cm,height=7.5cm]{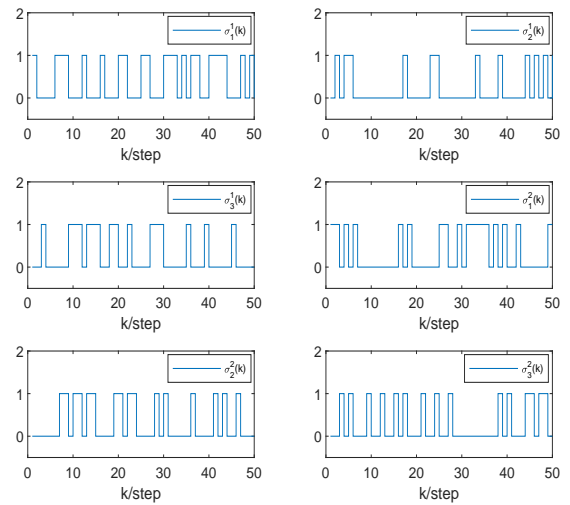}
	\caption{The sequences of the variables $\sigma^i_{\kappa_i}(k)(i=1,2;\kappa_i=1,2,3)$.}
	\label{fig4}
\end{figure}

In order to show the effectiveness of the proposed nonlinear fusion estimation method that is summarized in Algorithm 1, the vehicle's motion trajectory and the fusion estimation trajectory are depicted in Fig. \ref{fig3} (a). It is seen from this subfigure that the introduced ETS \eqref{ea12}-\eqref{eaa12} and DRS \eqref{re1}-\eqref{ea5} are such that the proposed nonlinear DFE algorithm can track the vehicle's motion trajectory well. Moreover, the estimation performance is evaluated in terms of the root mean square error (RMSE), and the RMESs of the LREs \eqref{eq8} and the DFE \eqref{eq10} with 100 Monte Carlo runs are shown in Fig. \ref{fig3} (b). It can be seen that the estimation error of the nonlinear fusion estimator is less than those of each local estimator, which implies that the estimation precision can be further improved by the developed nonlinear fusion estimation algorithm under the communication strategy presented in Sec. II-A.
However, it should be pointed out that the superiority of fusion estimation over local estimation in Fig. \ref{fig3} (b) is not as pronounced as it would be without resource constraints in Fig. 6 of \cite{q2}, this is mainly because ETS and DRS reduce the amount of communication as well as redundant information. At the same time, the triggering status of two group measurements is plotted in Fig. \ref{fig3} (c), where the value ``1'' indicates that the raw measurements can be further dimensionality reduced; otherwise, the corresponding remote estimator cannot receive any measurements from the sensor node. As shown in this subfigure, there are some untriggered instants in the corresponding time interval, which implies that the measurements can be effectively reduced based on the proposed ETS to meet the limited resources. On the other hand, the sequences of the decision variables $\sigma^i_{\kappa_i}(k)$ in \eqref{SA2} are plotted in Fig. \ref{fig4}. It can be seen that $\varsigma^i_m=\sum\nolimits^3_{\kappa_i=1}\sigma^i_{\kappa_i}(k)=2$ and $\varsigma_m=4$ satisfy the bandwidth constraint \eqref{re1}, and then the corresponding dimensionality matrix $\Theta^i_m(k)$ in \eqref{SA2} is also determined. Therefore, it can be summarized from Fig. \ref{fig3} (a-c) and Fig. \ref{fig4} that Algorithm 1 is effective in the reduction of information transmission while preserving satisfactory estimation performance.

\begin{figure}[!htb]
	\setlength{\abovecaptionskip}{0cm}
	\centering
	\includegraphics[width=9cm,height=8cm]{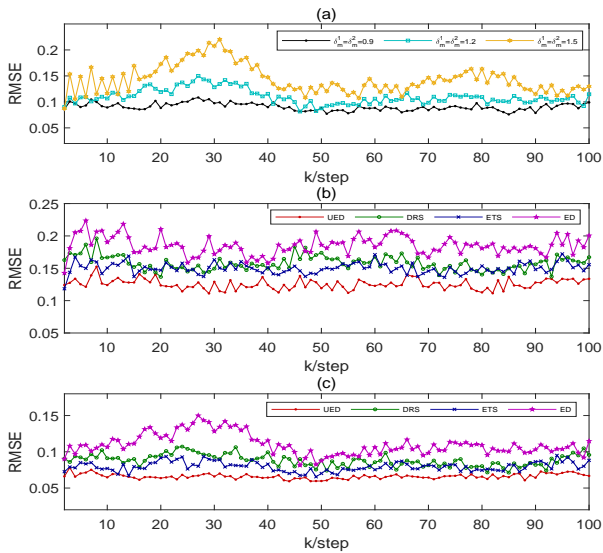}
	\caption{(a) The effect of fusion estimation under different triggering thresholds; (b) The effect of fusion estimation under different communication strategies with Gaussian noises; (c) The effect of fusion estimation under different communication strategies with bounded noises.}
	\label{fig5}
\end{figure}

In particular, the comparison of fusion estimation performance for different event-triggered thresholds $\delta^i_m$ is depicted in Fig. \ref{fig5} (a), respectively. It is shown that the smaller the threshold is selected, the higher the estimation precision is given. This is because the estimation performance under ETS depends on the choice of event-triggered thresholds, that is the smaller threshold can provide more measurements transmitted to the corresponding remote estimators. In this sense, the threshold cannot be selected too small due to the limitation of the resources. It gives a requirement to select the appropriate thresholds to minimize the communication traffic while keeping a satisfactory estimation precision. Whereafter, Fig. \ref{fig5} (b) shows the fusion estimation performance under different communication strategies with Gaussian noises. It is obvious that the estimation accuracy without resource constraints (the red line) is the highest compared with several other communication strategies, which is because the local estimator provides more redundant information to the FC. If only DRS was used to deal with the bandwidth constraints, it would also cause a degradation in estimation performance (the green line). This is because some communication bandwidth was abandoned to transmit the measurement information, while some information will still be transmitted to the remote estimator each time. Moreover, if only using ETS to meet the resource constraints, the estimation performance is also slightly degraded relative to without resource constraints (the blue line). This is because the ETS will cause the entire measurement information to be discarded at a particular time. In comparison, when considering ETS and DRS in a unified framework, the estimation performance is the worst among several communication strategies (the magenta line), but more communication traffic can be reduced to solve the estimation problem in NMFSs under resource constraints, and the estimation performance is not severely degraded either. Meanwhile, the estimation performance under different communication strategies with bounded noises \eqref{s3} is shown in Fig. \ref{fig5} (c), which has a similar effect to the Gaussian noise case. Since the compensation model \eqref{ea14} is proposed in this paper, the estimation performance of the other communication strategies is not seriously degraded as compared without resource constraints.

\begin{figure}[!htb]
	\setlength{\abovecaptionskip}{0cm}
	\centering
	\includegraphics[width=9cm,height=9.5cm]{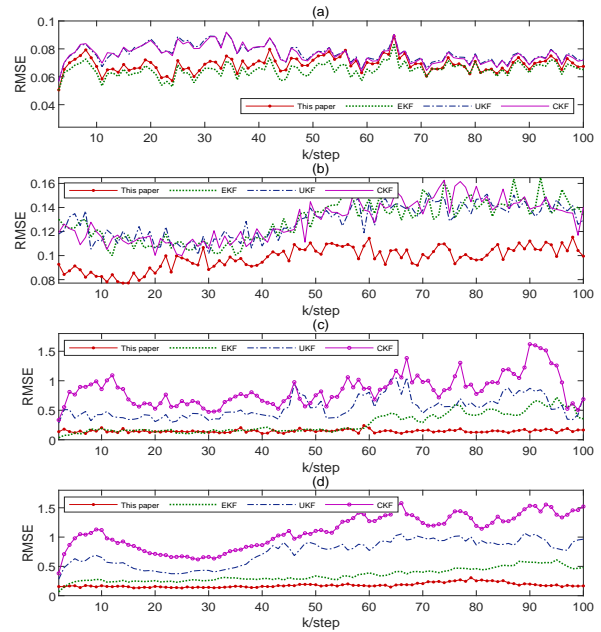}
	\caption{(a) The comparison of estimation performance among the proposed methods, EKF \cite{b26}, UKF \cite{b27} and CKF \cite{b28} under Gaussian noise without resource constraints; (b) The comparison of estimation performance among those methods under bounded noise without resource constraints; (c) The comparison of estimation performance among those methods under Gaussian noise with resource constraints; (d) The comparison of estimation performance among those methods under bounded noise with resource constraints.}
	\label{fig6}
\end{figure}

When considering different communication strategies and noise conditions, the comparison of estimation performance between the proposed method and the EKF \cite{b26}, UKF \cite{b27}, and CKF \cite{b28} methods, which are classical methods to deal with nonlinear estimation problems, is presented in Fig. \ref{fig6}. Specifically, Fig. \ref{fig6} (a) shows the estimation effect of these methods under Gaussian noise without resource constraints, respectively, where the noise covariances are set as $Q_w=\mathrm{diag}\{1,1,1\} \times 10^{-3}$, $Q_v=\mathrm{diag}\{1,1,1\} \times 10^{-2}$. It is seen from this subfigure that the EKF has the best estimation accuracy in this example, the UKF and CKF have similar estimation results, and the proposed estimator \eqref{eq8} in this paper also has ideal estimation performance. However, when the statistical information of the system noises cannot be obtained accurately (such as \eqref{s3}), the estimation performance of these methods is shown in Fig. \ref{fig6} (b). It can be seen that the RMSE of the proposed method is the lowest, and the RMSEs of the EKF, UKF, and CKF methods are very similar, which indicates that the estimation accuracy of the proposed method is better than other methods in the case of unknown noise statistical information. Notice that, this paper focuses on the estimation problem of NMFSs under resource constraints. Thus, the estimation performance comparisons of these methods under Gaussian noises and bounded noises \eqref{s3} are depicted in Fig. \ref{fig6} (c) and Fig. \ref{fig6} (d), respectively. It can be seen from the subfigures that the proposed nonlinear estimation method based on ETS \eqref{ea12}-\eqref{eaa12} and DRS \eqref{re1}-\eqref{ea5} has better estimation accuracy in both noise cases. In fact, the dimensionality reduction matrix may cause problems with matrix factorization or the propagation of non-positive definite covariance matrices in UKF and CKF methods at a particular time. Moreover, these classical nonlinear estimation methods and various extension methods are usually designed to deal with Gaussian noises with known covariances, while the bounded noises with unknown statistical information are addressed in this paper for designing nonlinear estimators. Therefore, in the case of bounded noises without accurate covariances, the designed nonlinear estimator has better estimation precision.

\subsection{ETS and DRS Based on Local Nonlinear Estimation}
When the S-FC channel is under limited resources, the system state \eqref{s1} can be estimated by implementing Algorithm 2. Then, the triggering thresholds in ETS \eqref{eb9}-(\ref{ebb9}) are set as $\delta^1_s=\delta^2_s=1.0$. Additionally, consider that there are two components of the local estimate $\hat{x}^s_i(k)$ that might be transmitted to the FC in this case, then $\mathfrak{h}^1_s=\mathfrak{h}^2_s=3$. Therefore, $\Theta^i_s(k)$ has a similar form as $\Theta^i_m(k)$ in \eqref{SA2}. Since the $\Theta^i_s(k)$ is decided by $\sigma^i_{\hbar_i}(k)$, and the stochastic process $\{\sigma^i_{\hbar_i}(k)\}$ in \eqref{eb17} obeys i.i.d.. Then, the selection probabilities are given as $\pi^1_1=0.3, \pi^1_2=0.2, \pi^1_3=0.5, \pi^2_1=0.3, \pi^2_2=0.3, \pi^2_3=0.4$, which satisfies $\sum^3_{\hbar_i=1}\pi^i_{\hbar_i}=1$. Moreover, the sate-dependent matrices $L^f_1=L^f_2=\mathrm{diag}\{0.03,0.01,0.02\}$, $L^g_1=L^g_2=\mathrm{diag}\{0.03,0.03,0.02\}$ and $\alpha^1_s=\alpha^2_s=1$.

\begin{figure}[!htb]
	\setlength{\abovecaptionskip}{0cm}
	\centering
	\includegraphics[width=9cm,height=7cm]{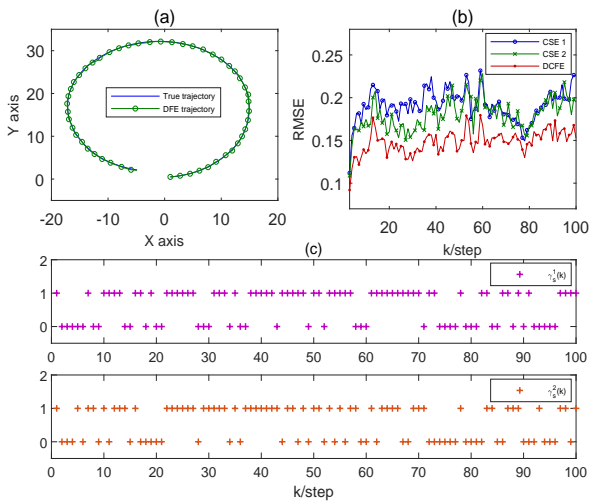}
	\caption{(a) The vehicle's true motion trajectory and fusion estimation trajectory; (b) The estimation performance comparison between the CSEs and DCFE; (c) The triggering status of two estimators.}
	\label{fig7}
\end{figure}

\begin{figure}[!htb]
	\setlength{\abovecaptionskip}{0cm}
	\centering
	\includegraphics[width=9cm,height=7.5cm]{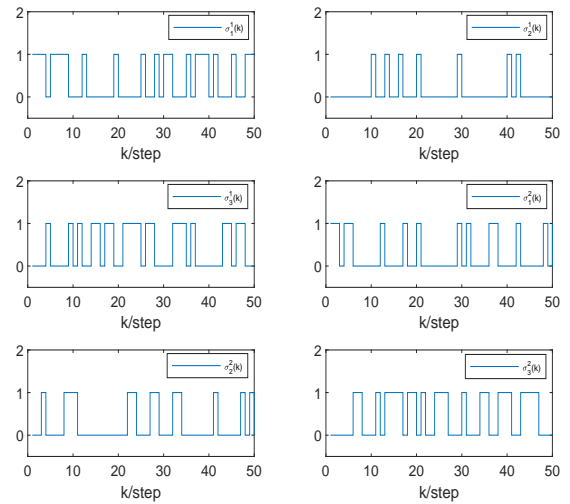}
	\caption{The sequences of the variables $\sigma^i_{\hbar_i}(k)(i=1,2;\hbar_i=1,2,3)$.}
	\label{fig8}
\end{figure}

To show the effectiveness of the proposed nonlinear fusion estimation Algorithm 2, the actual vehicle's trajectory and the DCFE trajectory are presented in Fig. \ref{fig7} (a), which shows that the proposed nonlinear compensation fusion estimator can get the vehicle's position under incomplete information. Then, the RMESs of the CSEs \eqref{eb12} and the DCFE \eqref{eb13} with 100 Monte Carlo runs are shown in Fig. \ref{fig7} (b), respectively. It can be seen from the subfigure that even under the ETS \eqref{eb9}-\eqref{ebb9} and DRS \eqref{eb3}-\eqref{eb15} presented in Sec. II-B, the performance of the DCFE is still better than that of each CSE under the action of the compensation strategy \eqref{eb12}. Moreover, the triggering status of two smart sensors is plotted in Fig. \ref{fig7} (c), where it can be seen that some untriggered instants indicate the communication traffic of the local estimation is really reduced. Meanwhile, based on the selection probability $\pi^i_{\hbar_i}$, the sequences of the variables $\sigma^i_{\hbar_i}(k)$ are plotted in Fig. \ref{fig8}, which satisfy $\varsigma^i_s=\sum\nolimits^3_{\hbar_i=1}\sigma^i_{\hbar_i}(k)=2$, and then the corresponding dimensionality matrix $\Theta^i_s(k)$ can also be determined at each time. Similarly, it is summarized from Fig. \ref{fig7} (a-c) and Fig. \ref{fig8} that Algorithm 2 provides an effective scheme to lighten the communication burden while keeping a certain estimation accuracy.

\begin{figure}[!htb]
	\setlength{\abovecaptionskip}{0cm}
	\centering
	\includegraphics[width=9cm,height=8cm]{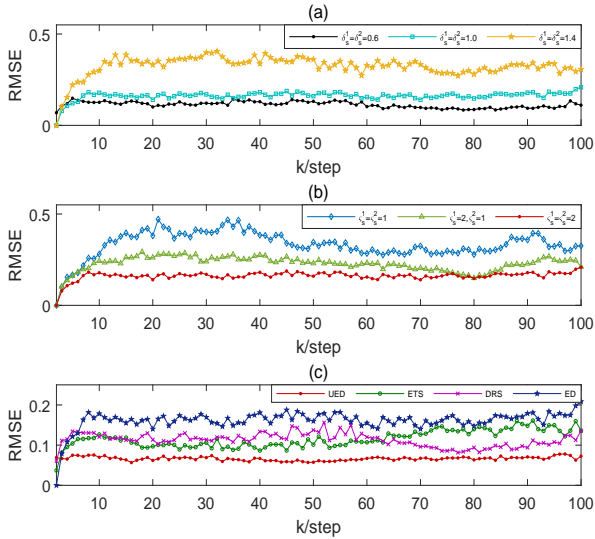}
	\caption{(a) The effect of fusion estimation under different triggering thresholds; (b) The effect of fusion estimation under different bandwidth constraints; (c) The effect of fusion estimation under different communication strategies.}
	\label{fig9}
\end{figure}

Furthermore, the comparison of the fusion estimation performance for different communication strategies is presented in Fig. \ref{fig9}. Firstly, the RMESs of DCFE for different event-triggered thresholds $\delta^i_s(i=1,2)$ is depicted in Fig. \ref{fig9} (a), which shows that the choice of thresholds can really affect the estimation performance, and the DCFE with small triggering thresholds has a higher estimation accuracy than that of other larger triggering thresholds. This is because of the smaller threshold given in (\ref{ebb9}), the more accurate local estimates might be sent to the FC, while the resource constraints may not be settled very well. Thus, the event-triggered thresholds give a trade-off between the estimation performance and the communication resources. Whereafter, Fig. \ref{fig9} (b) shows the RMESs for different dimensionality reduction states (i.e. $\varsigma^1_s=\varsigma^2_s=1$, $\varsigma^1_s=2, \varsigma^2_s=1$, $\varsigma^1_s=\varsigma^2_s=2$, respectively) in this example. It can be seen from this subfigure that the more components of the local estimate transmitted to the FC has a better estimation accuracy under the same triggering thresholds. Indeed, this is because the capacity of the bandwidth channel limits the transmission of local estimates. Lastly, Fig. \ref{fig9} (c) shows the RMSEs under different resource constraint cases. Similarly, the NMFSs without resource constraints have the best fusion estimation accuracy. As the analysis in the above theoretical section, only using ETS or DRS will reduce the burden of communication resources to a certain extent, and the estimation will be degraded as well. In comparison, though the ETS and DRS in the unified framework considered in this paper have the worst estimation accuracy among these communication strategies, they can reduce more communication traffic to meet the requirements of the resource constraints and the fusion estimation performance is still maintained to an acceptable extent.

\section{Conclusion}
In this paper, two different communication frameworks for the nonlinear NMFSs, where the S-RE channel and the S-FC channel were subject to resource constraints, respectively, have been considered. Specifically, an ETS and a DRS were utilized to alleviate the communication burden, which could meet the finite communication resources. Meanwhile, in order to preserve a certain estimation performance, a unified compensation model was designed to restructure the reduced information. Then, the nonlinear local/fusion estimators were proposed based on the compensation information, and the uncertain parameters together with the state-dependent matrices were introduced to model the linearization errors. In this case, a bounded recursive optimization approach, which can dealt with the distributed fusion estimation problem for nonlinear NMFSs with unknown noise statistical information, was employed to solve the nonlinear estimator gains. Moreover, the proposed robust design approach can be such that the SEs of the designed compensation estimators were bounded as well. Finally, two simulation cases were presented to illustrate the effectiveness and advantages of the proposed methods.

Furthermore, the time-delay, asynchronous sampling, and out of order will cause an asynchronous fusion problem in the nonlinear networked fusion structure, which has also attracted significant attention. In particular, some neural networks \cite{r50}-\cite{r51} and robust methods \cite{r52,r53,r54} give us an inspring for future work dealing with unknown perturbations, linearization errors, and uncertainty problems in the field of asynchronous multi-sensor fusion systems. Therefore, how to design stable estimators for nonlinear asynchronous NMFSs will be one of our future tasks.

\appendices
\section{Proof of the Theorem 1}
By using the definition of some complex variables in TABLE II of Appendix C, the local estimation error $\tilde{x}^m_i(k)$ in (\ref{eq16}) can be rewritten as:
\begin{equation}\label{eq19}
\tilde{x}^m_i(k)={{A}}_{\mathrm{K}}^i(k)\tilde{{x}}^m_i(k-1)+{\Gamma}_{\mathrm{K}}^i(k){{w}}(k-1)
+{D}_{\mathrm{K}}^{i}(k){v}_i(k)
\end{equation}
Then, a performance index \cite{b34} is introduced as follows:
\begin{equation}
\begin{aligned}\label{eq20}
\mathrm{J}_i^m(k)\triangleq &\;(\tilde{x}^m_i(k))^{\mathrm{T}}\tilde{x}^m_i(k)-(\tilde{x}^m_i(k-1))^{\mathrm{T}}
\Psi_i(k)\tilde{x}^m_i(k-1)\\
-&({w}(k-1))^{\mathrm{T}}\Phi_i(k){w}(k-1)-{v}^{\mathrm{T}}_i(k)\Upsilon_i(k){v}_i(k)
\end{aligned}\end{equation}
where $\Psi_i(k)$, $\Phi_i(k)$ and $\Upsilon_i(k)$ are unknown positive definite matrices. In fact, it can be deduced from the above index that $\mathrm{J}^m_i(k)<0$ can construct an upper bound of the SE of the LRE \eqref{eq8}. Then, from \eqref{eq19}-\eqref{eq20} and Schur complement lemma \cite{b35}, the $\mathrm{J}^m_i(k)<0$ is equivalent to the following inequality:
\begin{equation}\label{ap1}
\left[
\begin{array}{cccc}
-I & {{A}}_\mathrm{K}^i(k) & {\Gamma}_\mathrm{K}^i(k) & {D}_{\mathrm{K}}^{i}(k)\\
* & -\Psi_i(k) & 0 & 0 \\
* & * & -\Phi_i(k) & 0 \\
* & * & * & -\Upsilon_i(k)
\end{array}
\right]<0
\end{equation}
Since the above matrix inequality contains several uncertain matrices that are introduced in linearized process \eqref{eq13}, then the inequality \eqref{ap1} can be converted to the first three inequalities in the optimization problem \eqref{eq18} by using the Lemma 1. Furthermore, the bounded stability condition of the designed LRE \eqref{eq8} can be obtained from the similar analysis of \cite[Th.1]{q2}. That is, the inequality (\ref{ee19}) is held when the last two inequalities in \eqref{eq18} are held. In this case, the upper bound of the SE of each LRE was scaled in this paper, and then it follows from $\mathrm{J}^m_i(k)<0$ that
\begin{equation}\begin{aligned}\label{eq29}
(\tilde{x}^m_i(k))^{\mathrm{T}}\tilde{x}^m_i(k)&<\zeta_i(k)(\tilde{x}^m_i(k-1))^{\mathrm{T}}
\tilde{x}^m_i(k-1) \\
&+\lambda_{\max}\left(\begin{bmatrix}
                  w(k-1) \\
                  v_i(k)
                \end{bmatrix}^{\mathrm{T}}
                \begin{bmatrix}
                  w(k-1) \\
                  v_i(k)
                \end{bmatrix}\right)\\
                &\times({\mathrm{Tr}}\{\Phi_i(k)\}+{\mathrm{Tr}}\{\Upsilon_i(k)\})\\
\end{aligned}\end{equation}
where $\zeta_i(k)$ has been defined in (\ref{eq18}). Notice that, in order to minimize the upper bound of the $(\tilde{x}^m_i(k))^{\mathrm{T}}\tilde{x}^m_i(k)$, $``\min~({\mathrm{Tr}}\left\{\Sigma_i(k)\right\}+{\mathrm{Tr}}\left\{\Upsilon_i(k)\right\})"$ was selected as the optimization objective to determine the LRE gain $\mathrm{K}^m_i(k)$, then the optimization problem \eqref{eq18} in terms of linear matrix inequalities (LMIs) was established.

Next, the distributed fusion matrix $\mathrm{W}^m_i(k)$ will be determined by the following analysis.
Let $\tilde{{x}}_m(k) \triangleq {x}(k)-\hat{{x}}_m(k)$ denote the fusion estimation error, it follows from (\ref{eq1}) and (\ref{eq10}) that
\begin{equation}\label{eq32}
\tilde{{x}}_m(k)=\sum\limits^L_{i=1}\mathrm{W}^m_i(k)\tilde{x}^m_i(k)
\end{equation}
Combining (\ref{eq19}) and (\ref{eq32}), the fusion error system can be formulated as
\begin{equation}\label{eq33}
\begin{aligned}
\tilde{x}_m(k)=&\;\mathrm{W}_m(k) \left[
{A}^m_\mathrm{K}(k)\tilde{x}_{L}^M(k-1) \right.\\
&\left. +{\Gamma}^m_{\mathrm{K}}(k){w}(k-1)+D^m_{\mathrm{K}}(k){v}(k)\right]
\end{aligned}\end{equation}
where $\tilde{x}_{L}^M(k)\triangleq {\mathrm{col}}\{\tilde{x}^m_1(k),\ldots,\tilde{x}^m_L(k)\}$, ${v}(k)\triangleq{\mathrm{col}}\{v_1(k),$ $\ldots,v_L(k)\}$, while ${A}_\mathrm{K}^m(k)$, ${\Gamma}_{\mathrm{K}}^m(k)$, $D_{\mathrm{K}}^m(k)$ and $\mathrm{W}_m(k)$ have been defined in TABLE II.

Similarly, by introducing some unknown matrices $\Psi(k)>0, \Phi(k)>0, \Upsilon(k)>0, \Psi_1(k), \Psi_2(k)$ and $\Phi_1(k)$ to construct an upper bound on the MSE of the DFE \eqref{eq10}, one obtains
\begin{equation}\label{eq34}
\begin{aligned}
\tilde{x}^{\mathrm{T}}_m(k)\tilde{x}_m(k) <  \begin{bmatrix}
                                              \tilde{x}_L^M(k-1) \\
                                              {w}(k-1) \\
                                              {v}(k)
                                            \end{bmatrix}^{\mathrm{T}} \Delta(k)
                                            \begin{bmatrix}
                                              \tilde{x}_L^M(k-1) \\
                                              {w}(k-1) \\
                                              {v}(k)
                                            \end{bmatrix}
\end{aligned}
\end{equation}
where
\begin{equation}\label{ap2}
 \Delta(k) \triangleq \left [\begin{array}{ccc} \Psi(k) & \Psi_1(k) & \Psi_2(k) \\  {*} & \Phi(k) & \Phi_1(k) \\ * & * & \Upsilon(k)\end{array}\right]
\end{equation}
Then, using a derivation process similar to that in \cite[Th.2]{q2}, the inequality \eqref{eq34} can be converted to the LMIs in \eqref{eq30}. Moreover, it is seen from Sec.II-A that minimizing the constructed upper bound of $\tilde{x}^{\mathrm{T}}_m(k)\tilde{x}_m(k)$ is one of the aims of this paper. Therefore, $``\min~({\mathrm{Tr}}\left\{\Psi(k)\right\}+{\mathrm{Tr}}\left\{\Phi(k)\right\}+{\mathrm{Tr}}\left\{\Upsilon(k)\right\})"$
is chosen as the optimization objective to construct an optimization problem to calculate distributed weighting fusion matrices $\{\mathrm{W}^m_i(k)\left|\right.\sum^L_{i=1}\mathrm{W}^m_i(k)=I, i\in\mathfrak{L}\}$.
The detailed derivation is omitted here. $\hfill\blacksquare$

\section{Proof of the Theorem 2}
Motivated the stability analysis of the compensation estimators in \cite[Th.2]{qd1}, the stability condition of the CSE \eqref{eb12} and the DCFE \eqref{eb13} are analyzed as follows.

Firstly, it follows from \eqref{ec1} and (\ref{ec3}) that
\begin{equation}\begin{aligned}\label{ec14}
\tilde{x}^c_i(k)=&\;(I-\gamma^i_s(k)\Theta^i_s(k))(A^i_c(k-1)+L^i_c(k)P^i_c(k))\\
&\times\tilde{x}_i^c(k-1)+\varpi_i(k)
\end{aligned}\end{equation}
where $\varpi_i(k)\triangleq \gamma^i_s(k)\Theta^i_s(k)\tilde{x}_i^s(k)+(I-\gamma^i_s(k)\Theta^i_s(k))\Gamma(k-1)w(k-1)$.
Then, it is derived from above equation that
\begin{equation}\label{ec15}
\begin{aligned}
\mathbb{E}\{\tilde{x}_i^c(k)\}=&\;\Theta_I^i(A^i_c(k-1)+L^i_c(k)P^i_c(k))\mathbb{E}\{\tilde{x}^c_i(k-1)\}\\
& +\mathbb{E}\{\varpi_i(k)\}
\end{aligned}
\end{equation}
where $\Theta_I^i=\mathbb{E}\{I-\gamma^i_s\Theta^i_s(k)\}$. Since the SE of the $\tilde{x}^s_i(k)$ is bounded from the analysis in \cite[Th.1]{q2}, and it is deduced from (\ref{eb15}) that $0<\|\Theta_I^i\|_2<1$, thus $\mathbb{E}\{\varpi(k)_i\}$ in (\ref{ec15}) is bounded as well. In this case, when considering the $\mathbb{E}\{\tilde{x}^c_i(k)\}$ is norm bounded, i.e.,
\begin{equation}\label{ec19}
\lim\limits_{k\rightarrow \infty}\|\mathbb{E}\{\tilde{x}^c_i(k)\}\|_2<\mathcal{M}^i_c
\end{equation}
where $\mathcal{M}^i_c$ is a positive scalar, then the following condition should be held:
\begin{equation}\label{ec16}
\begin{aligned}
\|\Theta_I^i(A^i_c(k-1)+L^i_c(k)P^i_c(k))\|_2<1
\end{aligned}
\end{equation}
Then, by the means of Schur complement Lemma \cite{b35} and Lemma 1, \eqref{ec16} holds is equivalent to (\ref{ec18}) holds.

Furthermore, the (\ref{ec14}) can also be rewritten as
\begin{equation}\begin{aligned}\label{ec20}
\tilde{x}^c_i(k+1)
=\Pi^i_c(k,N_i)\tilde{x}^c_i(k-N_i)+\vartheta_i(k)
\end{aligned}\end{equation}
where
\begin{equation} \label{ec21}
\left\{
\begin{aligned}
&\Pi^i_c(k,N_i)\triangleq \mathop{\prod}^{N_i}_{\iota=0}(I-\gamma^i_s(k-\iota+1)\Theta^i_s(k-\iota+1))\\
&\;\;\;\;\;\;\;\;\;\;\;\;\;\; \times (A_c^i(k-\iota)+L_c^i(k-\iota+1)P_c^i(k-\iota+1)) \\
&\vartheta_i(k) \triangleq \sum^{N_i}_{\iota=0}\left\{\Pi^i_c(k,\iota-1)\{\gamma^i_s(k-\iota+1)\Theta^i_s(k-\iota+1)\right. \\
&\qquad \times \tilde{x}_i^s(k-\iota+1)+(I-\gamma^i_s(k-\iota+1)\Theta^i_s(k-\iota+1))\\
&\qquad \times \Gamma(k-\iota)w(k-\iota)\}\}
\end{aligned}\right.
\end{equation}
Since the stochastic process  $\{\sigma^i_{\hbar_i}(k)\}$ is assumed to be i.i.d, and from (\ref{ec20}), one has
\begin{equation}\begin{aligned}\label{ec26}
&\mathbb{E}\{(\tilde{x}^c_i(k+1))^\mathrm{T}\tilde{x}^c_i(k+1)\}\\
&=\mathbb{E}\{(\tilde{x}^c_i(k-N_i))^\mathrm{T}\mathcal{I}_c^i\tilde{x}^c_i(k-N_i)\}+\mathcal{O}^i_{c}(k)\\
&\leq \lambda_{\max}(\mathbb{E}\{\mathcal{I}_c^i\})\mathbb{E}\{(\tilde{x}^c_i(k-N_i))^\mathrm{T}
\tilde{x}^c_i(k-N_i)\}+\mathcal{O}^i_{c}(k)
\end{aligned}\end{equation}
where $\mathcal{I}_c^i \triangleq (\Pi_c^i(k,N_i))^{\mathrm{T}}\Pi_c^i(k,N_i)$, $\mathcal{O}^i_{c}(k)\triangleq \mathbb{E}\{\vartheta^\mathrm{T}_i(k)\vartheta_i(k)\}+2\mathbb{E}\{(\tilde{x}^c_i(k-N_i))^\mathrm{T}\}
\mathbb{E}\{(\Pi^i_c(k,N_i))^\mathrm{T}\vartheta_i(k)\}$. Then, according to \eqref{ec21} yields that
\begin{equation}\begin{aligned}\label{ec23}
\mathcal{I}^i_c
=&\left(\mathop{\prod}^{k}_{\iota=k-N_i}(A_c^i(k-\iota)+L_c^i(k-\iota+1)P_c^i(k-\iota+1))^\mathrm{T} \right.\\
&\left. \times (I-\gamma^i_s(k-\iota+1)\Theta^i_s(k-\iota+1))^\mathrm{T}\right)\\
&\times \left(\mathop{\prod}^{k-N_i}_{\iota=0}(I-\gamma^i_s(k-\iota+1)\Theta^i_s(k-\iota+1)) \right. \\
&\times (A_c^i(k-\iota)+L_c^i(k-\iota+1)P_c^i(k-\iota+1)) \\
= &\;{\mathcal{R}_i}(k-N_i,{\mathcal{R}_i}(k-N_i-1,{\mathcal{R}_i}(\ldots,{\mathcal{R}_i}(k,I))))
\end{aligned}\end{equation}
where ${\mathcal{R}_i}(k,Q)\triangleq (A^i_c(k)+L^i_c(k+1)P^i_c(k+1))^\mathrm{T}(I-\gamma^i_s(k+1)\Theta^i_s(k+1))Q
(I-\gamma^i_s(k+1)\Theta^i_s(k+1))(A^i_c(k)+L^i_c(k+1)P^i_c(k+1))$. In this case, it is deduced from (\ref{eb17}) and (\ref{ec23}) that
\begin{equation}\begin{aligned}\label{ec25}
\mathbb{E}\{\mathcal{I}_c^i\}= \breve{\mathcal{R}_i}(k-N_i,{\mathcal{R}_i}(k-N_i-1,\breve{\mathcal{R}_i}(\ldots,\breve{\mathcal{R}_i}(k,I))))
\end{aligned}\end{equation}
where $\mathbb{E}\{{\mathcal{R}_i}(k,Q)\}=\breve{\mathcal{R}_i}(k,Q)$. Moreover, it is seen from (\ref{eq3}), (\ref{ec19}) and \eqref{ec21} that $\mathcal{O}_{c}^i(k)$ in \eqref{ec26} is bounded. Therefore, it can be concluded from \eqref{ec26} and \eqref{ec25} that
$\lim_{k\rightarrow \infty } \mathbb{E}\{(\tilde{x}^c_i(k))^\mathrm{T}\tilde{x}^c_i(k)\}$ is bounded when $\lambda_{\max}(\mathbb{E}\{\mathcal{I}_c^i\})<1$, this condition can be scaled to the inequality (\ref{ec28}) holds.

On the other hand, the distributed weighting fusion matrix $\mathrm{W}^s_i(k)$ will be determined in the following.
Specifically, define $\tilde{x}_s(k)\triangleq \mathrm{col}\{\tilde{x}^s_1(k),\ldots,\tilde{x}^s_L(k)\}$, $\tilde{x}^s_c(k)\triangleq \mathrm{col}\{\tilde{x}^c_1(k),\ldots,\tilde{x}^c_L(k)\}$ and $\tilde{x}^o_s(k)\triangleq \mathrm{col}\{\tilde{x}_s(k),\tilde{x}^s_c(k)\}$, then combining (\ref{ec1}) and (\ref{ec3}), one has
\begin{equation}\label{ec5}
\tilde{x}_o^s(k)={A}_o^s(k)\tilde{x}_o^s(k-1)+{\Gamma}_o^s(k)w(k-1)+{D}_o^s(k)v(k)
\end{equation}
where $v(k)\triangleq \mathrm{col}\{v_1(k),\ldots,v_L(k)\}$ and
\begin{equation}\label{bp1}
  \begin{aligned}
  &A_o^s(k)\triangleq \begin{bmatrix}{A}_L^s(k) & 0 \\ \Theta_{\gamma}(k){A}_L^s(k) & {A}_L^c(k)  \end{bmatrix} \\
  &\Gamma_o^s(k)\triangleq \begin{bmatrix}{\Gamma}_L^s(k) \\  {\Gamma}_L^c(k)  \end{bmatrix},
  D_o^s(k)\triangleq \begin{bmatrix}{D}_L^s(k) \\ \Theta_{\gamma}(k){D}_L^s(k)  \end{bmatrix}
  \end{aligned}
\end{equation}
where ${A}_L^s(k), {A}_L^c(k), {\Gamma}_L^s(k), {\Gamma}_L^c(k)$, and ${D}_L^s(k)$ are defined in TABLE III.

Subsequently, the fusion error system can be constructed by (\ref{eq1}), (\ref{eb13}) and (\ref{ec5}) that
\begin{equation} \label{ec7}
\begin{aligned}
&\tilde{x}_c(k)={\mathrm{W}}^o_s(k)\tilde{x}^o_s(k)
\end{aligned}
\end{equation}
where ${\mathrm{W}}^o_s(k)=[0\quad \mathrm{W}_s(k)]$. Then, the upper bounded of MSE of the $\tilde{x}_c(k)$ can be determined by $\mathbb{E}\{\tilde{x}^{\mathrm{T}}_c(k)\tilde{x}_c(k)\}<\lambda_{\max}(\mathrm{W}^{\mathrm{T}}_s(k)\mathrm{W}_s(k))
\mathbb{E}\{(\tilde{x}^s_c(k))^{\mathrm{T}}\tilde{x}^s_c(k)\}$. Thus, when the conditions \eqref{ec18} and \eqref{ec28} are satisfied, the MSE of the DCFE \eqref{eb13} is bounded. Moreover, by using a similar analysis as Theorem 1, the optimization problem (\ref{ecc1}) is established to calculate $\mathrm{W}^s_i(k)$, and the detailed proof is omitted here. $\hfill\blacksquare$

\section{~}
\renewcommand\arraystretch{1.3}
\begin{table}[htbp]
\centering
\caption{The definition of the formulas}
\begin{tabularx}{9cm}{lX}
\toprule
Symbol & Formula \\
\midrule
${A}_{f_i}^m(k)$ & $\left.\partial {f}({x}(k))/ \partial {x}(k)\right|_{{x}(k)=\hat{x}^m_i(k)}$ \\
${C}_{h_i}^m(k)$ & $\left.\partial {h}_i({x}(k))/\partial {x}(k)\right|_{x(k)=\hat{x}^{-}_{m_i}(k)}$ \\
$\mathrm{K}_{C_i}^{m}(k)$  & $I-{\gamma^i_m(k)}{\mathrm{K}}^m_i(k){\Theta^i_m(k)}{C}_{h_i}^m(k)$ \\
$\mathrm{{K}}_{\Theta_i}^{m}(k)$ & $\gamma_m^i(k)\mathrm{K}^m_i(k)\Theta_m^i(k)$ \\
$\mathrm{K}^{A}_{C_i}(k)$ & $\mathrm{K}_{C_i}^{m}(k){A}_{f}^i(k-1)$ \\
$\mathrm{K}^\Gamma_{C_i}(k)$ & $\mathrm{K}_{C_i}^{m}(k){\Gamma}(k-1)$ \\
${{A}}_{\mathrm{K}}^i(k)$ & $\mathrm{K}^{A}_{C_i}(k)+\mathrm{K}_{C_i}^{m}(k){M}_{f}^i(k){N}_{f}^i(k)
                              -\mathrm{K}_{\Theta_i}^m(k){M}_{h}^i(k){N}_{h}^i(k)$ \newline
                             $\times{A}_{f}^i(k-1)-\alpha^i_m(k)\mathrm{K}^m_{\Theta_i}(k){M}_{h}^i(k){N}_m^i(k)$  \\
${\Gamma}_{\mathrm{K}}^{i}(k)$ & $(\mathrm{K}_{C_i}^{m}(k)-\mathrm{K}^m_{\Theta_i}(k){M}_{h}^i(k){N}_{h}^i(k)){\Gamma}(k-1)$ \\
$D_{\mathrm{K}}^{i}(k)$ & $-\gamma_m^i(k)\mathrm{K}^m_i(k)\Theta_m^i(k)D_i(k)$ \\
$N_m^i(k)$  & $(\alpha^i_m(k))^{-1}{N}_{h}^i(k){M}_{f}^i(k){N}_{f}^i(k)$ \\
$\mathrm{K}^M_{C_i}(k)$ & $ \begin{bmatrix}(\mathrm{K}_{C_i}^{m}(k){M}_{f}^i(k))^\mathrm{T} & 0 & 0 & 0\end{bmatrix}^\mathrm{T}$ \\
$\mathrm{K}_{\Theta_i}^M(k)$ & $\begin{bmatrix}-(\mathrm{K}^m_{\Theta_i}(k){M}_{h}^i(k))^\mathrm{T} & 0 & 0 & 0\end{bmatrix}^\mathrm{T}$\\
$O_I^i(k)$ & $\begin{bmatrix} 0 & \epsilon_{1i}(k)I & 0 & 0\end{bmatrix}$\\
$O_{\alpha}^i(k)$ & $\begin{bmatrix} 0 & \epsilon_{3i}(k)\alpha^i_m(k)I & 0 & 0\end{bmatrix}$ \\
$O_A^i(k)$ & $\begin{bmatrix} 0 & \epsilon_{2i}(k){A}_{f_i}^m(k-1) & \epsilon_{2i}(k){\Gamma}(k-1) & 0\end{bmatrix}$ \\
${{A}}_{\mathrm{K}}^m(k)$ & ${\mathrm{diag}}\{{{A}}_{\mathrm{K}}^1(k),\ldots,{{A}}_{\mathrm{K}}^L(k)\}$ \\
${A}_F^m(k)$ & $\mathrm{diag}\{{A}_{f_1}^m(k),\ldots,{A}_{f_L}^m(k)\}$ \\
${\Gamma}_{\mathrm{K}}^m(k)$ & $\mathrm{col}\{{\Gamma}_{\mathrm{K}}^{1}(k),\ldots,{\Gamma}_{\mathrm{K}}^{L}(k)\}$ \\
${\Gamma}_L(k)$ & $\mathrm{col}\{{\Gamma}(k),\ldots,{\Gamma}(k)\}$ \\
$D_{\mathrm{K}}^m(k)$ & $\mathrm{diag}\{D_{\mathrm{K}}^1(k),\ldots,D^L_{\mathrm{K}}(k)\}$ \\
$\alpha_{L}^m(k)$ & $\mathrm{diag}\{\alpha^1_m(k)I,\ldots,\alpha_m^L(k)I\}$ \\
$\mathrm{K}^{M}_{F}(k)$ & $\mathrm{diag}\{\mathrm{K}_{C_1}^{m}(k){M}_{f}^1(k),\ldots,\mathrm{K}_{C_L}^{m}(k){M}_{f}^L(k)\}$ \\
$\mathrm{K}^M_{H}(k)$ & $\mathrm{diag}\{\mathrm{K}^m_{\Theta_1}(k)M_{h}^1(k),\ldots,\mathrm{K}^m_{\Theta_L}(k)M_{h}^L(k)\}$ \\
$\mathrm{K}^{\mathrm{W}}_A(k)$ & $\mathrm{W}_m(k)\mathrm{diag}\{\mathrm{{K}}^A_{C_1}(k),\ldots,\mathrm{K}^A_{C_L}(k)\}$ \\
$\mathrm{K}^{\mathrm{W}}_{\Gamma}(k)$ & $\mathrm{W}_m(k)\mathrm{diag}\{\mathrm{{K}}^{\Gamma}_{C_1}(k),\ldots,\mathrm{K}^{\Gamma}_{C_L}(k)\}$ \\
$\mathrm{K}^{\mathrm{W}}_{D}(k)$ & $\mathrm{W}_m(k)\mathrm{diag}\{D_{\mathrm{K}}^1(k),\ldots,D^L_{\mathrm{K}}(k)\}$\\
$O^m_I(k)$ & $\begin{bmatrix} 0 & \epsilon_{1}(k)I & 0 & 0\end{bmatrix}$ \\
$O^m_{\alpha}(k)$ & $\begin{bmatrix} 0 & \epsilon_{3}(k)\alpha_L^M(k) & 0 & 0\end{bmatrix}$ \\
$O^m_A(k)$ & $\begin{bmatrix} 0 & \epsilon_{2}(k){A}_{F}^M(k-1) & \epsilon_{2}(k){\Gamma}_{L}(k-1) & 0\end{bmatrix}$ \\
$\mathrm{K}^{\mathrm{W}}_F(k)$ & $\begin{bmatrix}(\mathrm{W}_m(k)\mathrm{K}^{M}_{F}(k))^\mathrm{T} & 0 & 0 & 0\end{bmatrix}^\mathrm{T}$ \\
$\mathrm{K}^{\mathrm{W}}_H(k)$ & $\begin{bmatrix}-(\mathrm{W}_m(k)\mathrm{K}^M_{H}(k))^\mathrm{T} & 0 & 0 & 0\end{bmatrix}^\mathrm{T}$ \\
$\mathrm{W}_m(k)$ & $\left[\mathrm{W}^m_1(k),\ldots,\mathrm{W}^m_{L-1}(k),I-\sum\nolimits^{L-1}_{i=1}\mathrm{W}^m_i(k)\right]$ \\
$\Delta_m^i(k)$ & $\frac{1}{3}\left[
  \begin{array}{cccc}
    -I \!&\! \mathrm{K}^{A}_{C_i}(k) \!&\! \mathrm{K}^\Gamma_{C_i}(k) \!&\! D^{i}_{\mathrm{{K}}}(k) \\
     * \!&\! -\Psi_i(k) \!&\! 0 \!&\! 0 \\
     * \!&\! * \!&\! -\Phi_i(k) \!&\! 0 \\
     * \!&\! * \!&\! * \!&\! -\Upsilon_i(k) \\
  \end{array}
\right]$ \\
$\Delta_m(k)$ & $\frac{1}{3}\left[
  \begin{array}{cccc}
    -I \!&\! \mathrm{K}^{\mathrm{W}}_{A}(k) \!&\! \mathrm{K}^{\mathrm{W}}_{\Gamma}(k) \!&\! \mathrm{K}^{\mathrm{W}}_{D}(k) \\
     * \!&\! -\Psi(k) \!&\! -\Psi_1(k) \!&\! -\Psi_2(k) \\
     * \!&\! * \!&\! -\Phi(k) \!&\! -\Phi_1(k) \\
     * \!&\! * \!&\! * \!&\! -\Upsilon(k) \\
  \end{array}
\right]$ \\
[1mm]
\bottomrule
\end{tabularx}
\end{table}

\begin{table}[htbp]
\centering
\caption{The definition of the formulas}
\begin{tabularx}{9cm}{lX}
\toprule
Symbol & Formula \\
\midrule
$A^s_{f_i}(k)$ & $\partial f(x(k))/\partial x(k)|_{x(k)=\hat{x}^s_i(k)}$\\
$C^s_{h_i}(k)$ & $\partial h_i(x(k))/\partial x(k)|_{x(k)=\hat{x}^-_i(k)}$ \\
$A_c^i(k)$ & $\partial f(x(k))/\partial x(k)|_{x(k)=\hat{x}_i^c(k)}$ \\
$\mathrm{K}^s_{C_i}(k)$ & $I-\mathrm{K}^s_i(k)C^s_{h_i}(k)$ \\
$P_s^i(k)$ & $({\alpha}_s^i(k))^{-1}P_h^i(k)L_f^i(k)P_f^i(k)$ \\
${A}_s^i(k)$ & $\mathrm{K}^s_{C_i}(k)A^s_{f_i}(k-1)+\mathrm{K}^s_{C_i}(k)L_f^i(k)P_f^i(k)
                -\mathrm{K}^s_i(k)L_h^i(k)$ \newline
               $\times P^i_h(k)A^s_{f_i}(k-1)-{\alpha}_s^i(k)\mathrm{K}^s_i(k)L_h^i(k)P_s^i(k)$  \\
${A}_{\theta}^i(k)$ & $(I-\gamma_s^i(k)\Theta_s^i(k))(A^i_c(k-1)+L^i_c(k)P^i_c(k))$ \\
${\Gamma}_s^i(k)$ & $(\mathrm{K}^s_{C_i}(k)-\mathrm{K}^s_i(k)L_h^i(k)P_h^i(k))\Gamma(k-1)$ \\
${\Gamma}_c^i(k)$ & $(I-\gamma_s^i(k)\Theta_s^i(k)\mathrm{K}^s_i(k)C^s_{h_i}(k)-\gamma_s^i(k)\Theta_s^i(k)\mathrm{K}^s_i(k)$ \newline $\times L_h^i(k)P_h^i(k))\Gamma(k-1)$  \\
${A}_L^s(k)$ & $\mathrm{diag}\{{A}_s^1(k),\ldots,{A}_s^L(k)\}$ \\
${A}_L^{\theta}(k)$ & $\mathrm{diag}\{{A}_{\theta}^1(k),\ldots,{A}_{\theta}^L(k)\}$ \\
$A^s_F(k)$ & $\mathrm{diag}\{A^s_{f_1}(k),\ldots, A^s_{f_L}(k)\}$ \\
$A^c_L(k)$ & $\mathrm{diag}\{A^1_{c}(k),\ldots, A^L_{c}(k)\}$ \\
${\Gamma}_L^s(k)$ & $\mathrm{col}\{{\Gamma}_s^1(k),\ldots,{\Gamma}_s^L(k)\}$ \\
${\Gamma}_L^c(k)$ & $\mathrm{col}\{{\Gamma}_c^1(k),\ldots,{\Gamma}_c^L(k)\}$ \\
$L^s_F(k)$ & $\mathrm{diag}\{L_f^1(k),\ldots, L_f^L(k)\}$ \\
$\alpha^s_L(k)$ & $\mathrm{diag}\{{\alpha}_s^1(k)I,\ldots, {\alpha}_s^L(k)I\}$ \\
$L_L^c(k)$ & $\mathrm{diag}\{L_c^1(k),\ldots, L_c^L(k)\}$ \\
$\mathrm{K}_s(k)$ & $\mathrm{diag}\{\mathrm{K}^s_{C_1}(k),\ldots, \mathrm{K}^s_{C_L}(k)\}$ \\
$\mathrm{K}^s_C(k)$ & $\mathrm{diag}\{\mathrm{K}^s_1(k)C^s_{h_1}(k),\ldots, \mathrm{K}^s_L(k)C^s_{h_L}(k)\}$ \\
$\mathrm{K}^s_L(k)$ & $\mathrm{diag}\{\mathrm{K}^s_1(k)L_h^1(k),\ldots, \mathrm{K}^s_L(k)L_h^L(k)\}$ \\
${D}_L^s(k)$ & $\mathrm{diag}\{-\mathrm{K}^s_1(k)D_1(k),\ldots,-\mathrm{K}^s_L(k)D_L(k)\}$ \\
$\Theta_{\gamma}(k)$ & $\mathrm{diag}\{\gamma_s^1(k)\Theta^1_s(k),\ldots, \gamma_s^L(k)\Theta_s^L(k)\}$ \\
$\Theta^I_{\gamma}(k)$ & $\mathrm{diag}\{I-\gamma_s^1(k)\Theta^1_s(k),\ldots, I-\gamma_s^L(k)\Theta_s^L(k)\}$ \\
${O}^s_I(k)$ & $\begin{bmatrix} 0 & \varrho_{1}(k)I & 0 & 0 & 0 \end{bmatrix}$ \\
${O}^s_A(k)$ & $\begin{bmatrix} 0 & \varrho_{2}(k){A}^s_{F}(k-1) & 0 & \varrho_{2}(k){\Gamma}_{L}(k-1) & 0 \end{bmatrix}$ \\
${O}^s_L(k)$ & $\begin{bmatrix} 0 & \varrho_{3}(k)\alpha^s_L(k) & 0 & 0 & 0 \end{bmatrix}$ \\
${O}^s_\mathrm{W}(k)$ & $\begin{bmatrix} 0 & 0 & \varrho_4(k)I & 0 & 0 \end{bmatrix}$ \\
$A^{\mathrm{W}}_C(k)$ & $\mathrm{W}_s(k)\begin{bmatrix} \Theta_{\gamma}(k)\mathrm{K}_s(k)A^s_F(k-1)& {\Theta}^I_{\gamma}(k)A_L^c(k-1)\end{bmatrix}$ \\
$\Gamma^{\mathrm{W}}_C(k)$ & $\mathrm{W}_s(k)(I-\Theta_{\gamma}(k)\mathrm{K}^s_C(k))\Gamma_L(k-1)$\\
$L_{\mathrm{K}}^{\mathrm{W}}(k)$ & $\begin{bmatrix}(\mathrm{W}_s(k)\Theta_{\gamma}(k)\mathrm{K}_s(k)L_F^s(k))^\mathrm{T} & 0 & 0 & 0 \end{bmatrix}^\mathrm{T}$ \\
$\mathrm{K}^\mathrm{W}_{\Theta}(k)$ & $\begin{bmatrix}-(\mathrm{W}_s(k)\Theta_{\gamma}(k)\mathrm{K}^s_L(k))^\mathrm{T} & 0 & 0 & 0 \end{bmatrix}^\mathrm{T}$ \\
${L}^\mathrm{W}_{\Theta}(k)$ & $\begin{bmatrix}-(\mathrm{W}_s(k)\Theta^I_{\gamma}(k)L_L^c(k))^\mathrm{T} & 0 & 0 & 0 \end{bmatrix}^\mathrm{T}$ \\
$\mathrm{W}_s(k)$ & $[\mathrm{W}^s_1(k),\ldots,\mathrm{W}^s_{L-1}(k),I-\sum\nolimits^{L-1}_{i=1}\mathrm{W}^s_i(k)]$ \\
$\Delta_c(k)$ & $\frac{1}{4}\left[
  \begin{array}{cccc}
    -I \!&\! A^{\mathrm{W}}_C(k) \!&\! \Gamma^{\mathrm{W}}_C(k) \!&\! \mathrm{W}_s(k)\Theta_{\gamma}(k)D_L^s(k) \\
     * \!&\! -{{\Xi}}(k) \!&\! -{{\Xi}_1}(k) \!&\! -{{\Xi}_2}(k) \\
     * \!&\! * \!&\! -{{\Lambda}}(k) \!&\! -{{\Lambda}_1}(k) \\
     * \!&\! * \!&\! * \!&\! -{{\Sigma}}(k) \\
  \end{array}
\right]$\\
[1mm]
\bottomrule
\end{tabularx}
\end{table}

% or


%\section*{Acknowledgment}


\ifCLASSOPTIONcaptionsoff
  \newpage
\fi


\begin{thebibliography}{1}


\bibitem{r1}
S. Sun, and Z. Deng, ``Multi-sensor optimal information fusion Kalman filter,''
\emph{Automatica}, vol. 40, no. 6, pp. 1017--1023, 2004.

\bibitem{r2}
X. Li, Y. Zhu, and C. Han,``Optimal linear estimation fusion-Part I: Unified fusion rules,''
\emph{IEEE Transactions on Information Theory}, vol. 49, no. 9, pp. 2192--2208, 2003.

\bibitem{r3}
H. Lin, and S. Sun, ``Optimal sequential fusion estimation with stochastic parameter perturbations, fading measurements, and correlated noises,''
\emph{IEEE Transactions on Signal Processing}, vol. 66, no.13  pp. 3571--3583, 2018.

\bibitem{m1}
Y. Xia, J. Shang, J. Chen, and G. Liu, ``Networked data fusion with packet losses and variable delays,''
\emph{IEEE Transactions on Systems, Man, and Cybernetics--Part B: Cybernetics}, vol. 39, no. 5, pp. 1107--1120, 2009.

%\bibitem{m2}
%B. Chen, W. Zhang, G. Hu and L. Yu, ``Networked fusion Kalman filtering with multiple uncertainties,''
%\emph{IEEE Transactions on Aerospace and Electronic Systems}, vol. 51, no. 3,  pp. 2232--2249, 2015.

%\bibitem{b5}
%B. Chen, W. Zhang, L. Yu, G. Hu and H. Song, ``Distributed fusion estimation with communication bandwidth constraints,''
%\emph{IEEE Transactions on Automatic Control}, vol. 60, no. 5, pp. 1398--1403, 2015.

\bibitem{b1}
Y. Wang, S. X. Ding, H. Ye, and G. Wang, `` A new fault detection scheme for networked control systems subject to uncertain time-varying delay,''
\emph{IEEE Transactions on Signal Processing}, vol. 56, no. 10, pp. 5258--5268, 2008.

\bibitem{b30}
X. Yang, W. Zhang, and Li Yu, ``A bank of decentralized extended information filters for target tracking in event-triggered WSNs,''
\emph{IEEE Transactions on Systems, Man, and Cybernetics: Systems}, vol. 50, no. 9, pp. 3281--3289, 2020.

%\bibitem{b2}
%J. Hu, Z. Wang, D. Chen and F. E. Alsaadi, ``Estimation, filtering and fusion for networked systems with network-induced phenomena: new progress and prospects,''
%\emph{Information Fusion}, vol. 31, pp. 65--75, 2016.


\bibitem{t1}
M. Ghosal, and Vittal Rao,
 ``Fusion of multirate measurements for nonlinear dynamic state estimation of the power systems,''
\emph{IEEE Transactions on Smart Grid}, vol. 10, no. 1, pp. 216--226, 2019.

%\bibitem{b3}
%S. Deshmukh, B. Natarajan and A. Pahwa, ``State estimation over a lossy network in spatially distributed cyber-physical systems,''
%\emph{IEEE Transactions on Signal Processing}, vol. 62, no. 15, pp. 3911--3923, 2014.

\bibitem{b7}
J. A. Roecker, and C. D. McGillem, ``Comparison of two-sensor tracking methods based on state vector fusion and measurement fusion,''
\emph{IEEE Transactions on Aerospace and Electronic Systems}, vol. 24, no. 4, pp. 447--449, 1988.

\bibitem{b6}
Z. Wang, and Y. Niu, ``Distributed estimation and filtering for sensor networks,''
\emph{International Journal of Systems Science}, vol. 42, no. 9, pp. 1421--1425, 2011.

\bibitem{qu1}
X. Shen, Y. Zhu, and Z. You, ``An efficient sensor quantization alagorithm for decentralized estimation fusion,''
\emph{Automatica}, vol. 47, pp. 1053--1059, 2011.

\bibitem{qu2}
B. Chen, L. Yu, W. Zhang, and H. Wang, ``Distributed H$_\infty$ fusion filtering with communication bandwidth constraints,''
\emph{Signal Processing}, vol. 96, pp. 284--289, 2014.

\bibitem{qu3}
S. Zhu, C. Chen, J. Xu, and et.al., ``Mitigating quantization effects on distributed sensor fusion: a least squares approach,''
\emph{IEEE Transactions on Signal Processing}, vol. 66, no. 13, pp. 3459--3474, 2018.

%\bibitem{qu4}
%B. Xiang, B. Chen, L. Yu, ``Distributed fusion estimation for unstable systems with quantized innovations,''
%\emph{IEEE Transactions on Systems, Man, and Cybernetics: Systems}, vol. 51, no. 10, pp. 6381--6387, 2021.

\bibitem{dr0}
B. Chen, D. W. C. Ho, G. Hu, and L. Yu, ``Delay-dependent distributed Kalman fusion estimation with dimensionality reduction in cyber-physical systems,''
\emph{IEEE Transactions on Cybernetics}, 2021, DOI: 10.1109/TCYB.2021.3119461.

\bibitem{dr1}
Y. Zhu, E. Song, J. Zhou, and Z. You, ``Optimal dimensionality reduction of sensor data in multisensor estimation fusion,''
\emph{IEEE Transactions on Signal Processing}, vol. 53, no. 5, pp. 1631--1639, 2005.

\bibitem{dr11}
I. D. Schizas, G. B. Giannakis, and Z. Q. Luo, ``Distributed estimation using reduced-dimensionality sensor observations,''
\emph{IEEE Transactions on Signal Processing}, vol. 55, no. 8, pp. 4284--4299, 2007.

\bibitem{tsp1}
H. Ma, Y. H. Yang, Y. Chen, K. J. R. Liu, and Q. Wang,
``Distributed state estimation with dimension reduction preprocessing,"
\emph{IEEE Transactions on Signal Processing}, vol. 62, no. 12, pp. 3098--3110, 2014.

\bibitem{q21}
M. Xu, Y. Zhang, D. Zhang, and B. Chen, ``Distributed robust dimensionality reduction fusion estimation under DoS attacks and uncertain covariances,''
\emph{IEEE Access}, vol. 9, pp. 10328--10337, 2021.

\bibitem{b9}
X. Zhang, Q. Han, and B. Zhang, ``An overview and deep investigation on sampled-data-based event-triggered control and filtering for networked systems,''
\emph{IEEE Transactions on Industrial Informatics}, vol. 13, no. 1, pp. 4--16, 2017.

\bibitem{b17}
T. Sebastian, and R. D'Andrea, ``Event-based state estimation with variance-based triggering,''
\emph{IEEE Transactions on Automatic Control}, vol. 59, no. 12, pp. 3266--3281, 2014.

\bibitem{tsp2}
J. Wu, Q. S. Jia, K. H. Johansson, and L. Shi,
``Event-based sensor data scheduling: trade-off between communication rate and estimation quality,"
\emph{IEEE Transactions on Automatic Control}, vol. 58, no. 4, pp. 1041--1046, 2013.

\bibitem{tsp3}
D. Shi, T. Chen, and L. Shi,
``An event-triggered approach to state estimation with multiple pointand
set-valued measurements,"
\emph{Automatica}, vol. 50, pp. 1641--1648, 2014.

\bibitem{tsp4}
D. Han, Y. Mo, J. Wu, and et.al.,
``Stochastic event-triggered sensor schedule for remote state estimation,"
\emph{IEEE Transactions on Automatic Control}, vol. 60, no. 10, pp. 2661--2675, 2015.

\bibitem{tsp5}
S. Weerakkody, Y. Mo, B. Sinopoli, D. Han, and L. Shi,
``Multi-sensor scheduling for state estimation with event-based, stochastic triggers,"
\emph{IEEE Transactions on Automatic Control}, vol. 61, no. 9, pp. 2695--2701, 2016.

\bibitem{tsp6}
A. Mohammadi, and K. N. Plataniotis,
``Event-based estimation with information-based triggering and adaptive update,"
\emph{IEEE Transactions on Signal Processing}, vol. 65, no. 18, pp. 4924--4939, 2017.

\bibitem{tsp61}
P. Cheng, S. He, V. Stojanovic, X. Luan, and Fei Liu,
``Fuzzy fault detection for Markov jump systems with partly accessible hidden information:
an event-triggered approach,"
\emph{ IEEE Transactions on Cybernetics},
2021, DOI: 10.1109/TCYB.2021.3050209.

\bibitem{b29}
C. Wen, Z. Wang, T. Geng, and F. E. Alsaadi, ``Event-based distributed recursive filtering for state-saturated systems with redundant channels,''
\emph{Information Fusion}, vol. 39, pp. 96--107, 2018.

\bibitem{b31}
L. Li, M. Niu, Y. Xia, H. Yang, and L. Yan, ``Event-triggered distributed fusion estimation with random transmission delays,''
\emph{Information Sciences}, vol. 475, pp. 67--81, 2019.

\bibitem{tsp7}
L. Li, M. Niu, Y. Xia, and H. Yang, ``Stochastic event-triggered distributed fusion estimation under jamming attacks,''
\emph{IEEE Transactions on Signal and Information Processing over Networks}, vol. 7, pp. 309--321, 2021.

\bibitem{tsp8}
W. Song, Z. Wang, J. Wang, F. E. Alsaadi, and J. Shan,
``Distributed auxiliary particle filtering with diffusion strategy for target tracking: a dynamic event-triggered approach,"
\emph{IEEE Transactions on Signal Processing}, vol. 69, pp. 328--340, 2021.

\bibitem{dr3}
B. Chen, W. Zhang, L. Yu, and et.al., ``Distributed fusion estimation with communication bandwidth constraints,''
\emph{IEEE Transactions on Automatic Control}, vol. 60, no. 5, pp. 1398--1403, 2015.

\bibitem{dr4}
B. Chen, D. W. C. Ho, W. Zhang, and L. Yu, ``Distributed dimensionality reduction fusion estimation for cyber-physical systems under DoS attacks,''
\emph{IEEE Transactions on Systems, Man, and Cybernetics: Systems}, vol. 49, no. 2, pp. 455--468, 2019.

\bibitem{dr2}
B. Chen, W. Zhang, and L. Yu, ``Distributed finite-horizon fusion Kalman filtering for bandwidth and energy constrained wireless sensor networks,''
\emph{IEEE Transactions on Signal Processing}, vol. 62, no. 4, pp. 797--812, 2014.

\bibitem{tsp9}
B. Chen, G. Hu, W. Zhang, and L. Yu, ``Distributed mixed $H_2/H_\infty$ fusion estimation with limited communication capacity,''
\emph{IEEE Transactions on Automatic Control}, vol. 61, no. 3, pp. 805--810, 2016.

\bibitem{qd1}
B. Chen, D. W. C. Ho, W. Zhang, and L. Yu, ``Networked fusion estimation with bounded nosies,''
\emph{IEEE Transactions on Automatic Control}, vol. 62, no. 10, pp. 5415--5421, 2017.

%\bibitem{b8}
%Q. Liu, Z. Wang, X. He and D. Zhou, ``A survey of event-based strategies on control and estimation,''
%\emph{Systems Science and Control Engineering}, vol. 2, no. 1, pp. 90--97, 2014.

%\bibitem{b10}
%L. Li, H. Yang, Y. Xia, H. Yang, ``Event-based distributed state estimation for linear systems under unknown input and false data injection attack,''
%Signal Processing, vol. 170, pp. 1--14, 2020.

\bibitem{tsp10}
J. Wu, X. Ren, D. Han, D. Shi, and L. Shi, ``Finite-horizon Gaussianity-preserving event-based sensor scheduling
in Kalman filter applications,''
\emph{Automatic}, vol. 72, pp. 100--107, 2016.

\bibitem{tsp11}
G. Battistelli, L. Chisci, and D. Selvi, ``A distributed Kalman filter with event-triggered communication and
guaranteed stability,''
\emph{Automatic}, vol. 93, pp. 75--82, 2018.

\bibitem{tsp12}
D. Yu, Y. Xia, L. Li, and D. Zhai, ``A distributed Kalman filter with event-triggered communication and
guaranteed stability,''
\emph{Automatic}, vol. 118, pp. 1--10, 2020.

\bibitem{b11}
X. He, C. Hu, Y. Hong, L. Shi, and H. Fang, ``Distributed Kalman filters with state equality constraints: time-based and event-triggered communications,''
\emph{IEEE Transactions on Automatic Control}, vol. 65, no. 1, pp. 28--43, 2020.

%\bibitem{b13}
%H. Yang, H. Li, Y. Xia, L. Li, ``Hierarchical fusion estimation for multi-sensor networked systems
%with transmission delays and packet dropouts,''
%Signal Processing, vol. 156, pp. 156--165, 2019.

%\bibitem{b16}
%H. Dong, X. Bu, N. Hou, et.al, ``Event-triggered distributed state estimation for a class of time-varying systems over sensor networks with redundant channels,''
%Information Fusion, vol. 36, pp. 243--250, 2017.

%\bibitem{b23}
%G. Wang, N. Li, Y. Zhang, ``An event based multi-sensor fusion algorithm with deadzone like measurements,''
%{Information Fusion}, vol. 42, pp. 111--118, 2018.

%\bibitem{b24}
%D. Han, Y. Mo, J. Wu, et.al, ``Stochastic event-triggered sensor schedule for remote state estimation,''
%\emph{IEEE Transactions on Automatic Control}, vol. 60, no. 10, pp. 2661--2675, 2015.

%\bibitem{b25}
%D. Shi, T. Chen, Ling Shi, ``On Set-Valued Kalman Filtering and Its Application to Event-Based State Estimation,''
%\emph{IEEE Transactions on Automatic Control}, vol. 60, no. 5, pp. 1275--1290, 2015.

%\bibitem{b25}
%A. Mohammadi, K. N. Plataniotis, ``Event-based estimation with information-based triggering and adaptive update,''
%{IEEE Transactions on Signal Processing}, vol. 65, no. 18, pp. 4924--4939, 2017.


%\bibitem{b31}
%H. Tan, B. Shen, Y. Liu, A. Alsaedi and B. Ahmad, ``Event-triggered multi-rate fusion estimation for uncertain system with stochastic nonlinearities and colored measurement noises,''
%\emph{Information Fusion}, vol. 36, pp. 313--320, 2017.


\bibitem{b32}
J. Mao, D. Ding, H. Dong, and X. Ge, ``Event-based distributed adaptive Kalman filtering with unknown covariance of process noises,''
\emph{IEEE Transactions on Systems, Man, and Cybernetics: Systems}, vol. 51, no. 10, pp. 6170--6182, 2021.

%\bibitem{b33}
%Z. Gu, X. Zhou, T. Zhang, F. Yang and M. Shen, ``Event-triggered filter design for nonlinear cyber--physical systems subject to deception attacks,''
%\emph{ISA Transactions}, vol. 104, pp. 130--137, 2020.

\bibitem{b18}
H. Yan, P. Li, H. Zhang, X. Zhan, and F. Yang, ``Event-triggered distributed fusion estimation of networked multisensor systems with limited information,''
\emph{IEEE Transactions on Systems, Man, and Cybernetics: Systems}, vol. 50, no. 12, pp. 5330--5337, 2020.

\bibitem{b311}
L. Li, M. Fan, Y. Xia, and C. Zhu, ``Recursive distributed fusion estimation for nonlinear stochastic systems with event-triggered feedback,''
\emph{Journal of the Franklin Institute}, vol. 358, vo. 14, pp. 7286--7307, 2021.

\bibitem{q2}
R. Wang, B. Chen, and  L. Yu, ``Distributed nonlinear fusion estimation without knowledge of noise statistical information: a robust design approach,''
\emph{IEEE Transactions on Aerospace and Electronic Systems}, vol. 57, no. 5, pp. 3107--3117, 2021.

\bibitem{q1}
D. Zhu, B. Chen, Z. Hong, and  L. Yu, ``Networked nonlinear fusion estimation under DoS attacks,''
\emph{IEEE Sensors Journal}, vol. 21, no. 5, pp. 7058--7066, 2021.

\bibitem{b36}
G. Calafiore, ``Reliable localization using set-valued nonlinear filters,''
\emph{IEEE Transactions on Systems Man and Cybernetics--Part A Systems and Humans}, vol. 35, no. 2, pp. 189--197, 2005.

\bibitem{b35}
S. P. Boyd, L. E. Chaoui, and V. Balakrishnan,
\emph{Linear Matrix Inequalities in System and Control Theory}, Philadelphia, PA: SIAM, 1994.

\bibitem{b34}
B. Chen, G. Hu, D. W. C. Ho, and L. Yu, ``A new approach to linear/nonlinear distributed fusion estimation problem,''
\emph{IEEE Transactions on Automatic Control}, vol. 64, no. 3, pp. 1301--1308, 2019.

\bibitem{b26}
K. Reif, S. G\"{u}nther, E. Yaz, and R. Unbehauen, ``Stochastic stability of the discrete-time extended Kalman filter,''
\emph{IEEE Transactions on Automatic Control}, vol. 44, no. 4, pp. 714--728, 1999.

\bibitem{b27}
S. J. Julier, J. K. Uhlmann, and H. F. Durrant-Whyte, ``A new method for the nonlinear transformation of means and covariances in filters and estimators,''
\emph{IEEE Transactions on Automatic Control}, vol. 45, no. 3, pp. 477--482, 2000.

\bibitem{b28}
I. Arasaratnam, and S. Haykin, ``Cubature Kalman filters,''
\emph{IEEE Transactions on Automatic Control}, vol. 54, no. 6, pp. 1254--1269, 2009.

\bibitem{r50}
J. Wang, Y. Alipouri, and B. Huang. ``Dual neural extended Kalman filtering approach for multirate sensor data fusion,''
\emph{IEEE Transactions on Instrumentation and Measurement},
vol. 70, pp. 1--9, 2020.

\bibitem{r51}
P. Cheng, H. Wang, V. Stojanovic, and et al. ``Asynchronous fault detection observer for 2-D Markov jump systems.'
\emph{Transactions on Cybernetics},
2021, DOI: 10.1109/TCYB.2021.3112699.

\bibitem{r52}
H. Tao, X. Li, W. Paszke, and et al. ``Robust PD-type iterative learning control for discrete systems with multiple time-delays subjected to polytopic uncertainty and restricted frequency-domain,''
\emph{Multidimensional Systems and Signal Processing},
vol. 32, no. 2, pp. 671--692, 2021.

\bibitem{r53}
H. Fang, G. Zhu, V. Stojanovic, and et al. ``Adaptive optimization algorithm for nonlinear Markov jump systems with partial unknown dynamics,''
\emph{International Journal of Robust and Nonlinear Control},
vol. 31, no. 6, pp. 2126--2140, 2021.

\bibitem{r54}
T. Iori, and T. Ohtsuka. ``Recursive elimination method in moving horizon estimation for a class of nonlinear systems and non-Gaussian noise,''
\emph{SICE Journal of Control, Measurement, and System Integration},
vol. 13, no. 6, pp. 282--290, 2020.

\end{thebibliography}
\end{document}